\documentclass[draftclsnofoot,onecolumn,12pt]{IEEEtran}
\usepackage{amsthm,amssymb,graphicx,multirow,amsmath,color,amsfonts,subcaption}%,ulem}
\usepackage[update,prepend]{epstopdf}
\usepackage[latin1]{inputenc}
\usepackage{tikz}
\usetikzlibrary{arrows,calc}		% Optional ticks libraries
\usepackage{bbm} % for \mathbbm{1}
\usepackage{pdfpages}
\usepackage{tabulary}
\usepackage{multirow}
\usepackage{comment}

\def\nba{{\mathbf{a}}}

\def\nbo{{\mathbf{o}}}
\def\nbp{{\mathbf{p}}}

\def\nbr{{\mathbf{r}}}

\def\nbu{{\mathbf{u}}}

\def\nbz{{\mathbf{z}}}
\def\nb0{{\mathbf{0}}}
\def\nb1{{\mathbf{1}}}
\def\nb1{{\mathbf{1}}}

\def\nbphi{\boldsymbol{\phi}}

\def\ncalA{{\mathcal{A}}}

\def\ncalD{{\mathcal{D}}}

\def\ncalF{{\mathcal{F}}}

\def\ncalI{{\mathcal{I}}}

\def\ncalR{{\mathcal{R}}}
\def\ncalS{{\mathcal{S}}}

\def\ncalV{{\mathcal{V}}}

\def\ncalZ{{\mathcal{Z}}}

\def\nbbE{{\mathbb{E}}}

\def\nbbP{{\mathbb{P}}}

\def\nbbR{{\mathbb{R}}}
\def\nbbS{{\mathbb{S}}}

\newtheorem{lemma}{Lemma}

\newtheorem{definition}{Definition}

\newtheorem{theorem}{Theorem}

\newtheorem{remark}{Remark}

\newtheorem{approximation}{Approximation}
	
\allowdisplaybreaks 

\begin{document}
\title{A Tractable Analysis of the Blind Spot Probability in Localization Networks under Correlated Blocking}
\author{Sundar Aditya,~\IEEEmembership{Student~Member,~IEEE}, Harpreet S. Dhillon,~\IEEEmembership{Member,~IEEE}, Andreas F. Molisch,~\IEEEmembership{Fellow,~IEEE} and Hatim Behairy
\thanks{Sundar Aditya and Andreas F. Molisch are with WiDeS, Ming Hsieh Dept. of Electrical Engineering, USC, Los Angeles, CA 90089, USA (Email:\{sundarad,molisch\}@usc.edu).}
\thanks{Harpreet S. Dhillon is with Wireless@VT, Bradley Dept. of Electrical and Computer Engineering, Virginia Tech, Blacksburg, VA 24061, USA (Email: hdhillon@vt.edu).}
\thanks{Hatim Behairy is with King Abdulaziz City for Science and Technology, P. O. Box 6086, Riyadh 11442, Saudi Arabia (Email: hbehairy@kacst.edu.sa) \hfill Revised: \today.}
\thanks{This work was supported by KACST under grant number 33-878. This paper was presented in part at the International Conference on Ubiquitous Wireless Broadband (ICUWB), 2016 held at Nanjing, China \cite{Adi_Har_Mol_2016}.}
}

\maketitle

\begin{abstract}
In localization applications, the line-of-sight between anchors and targets may be blocked by obstacles in the environment. A target that is \emph{invisible} (i.e., without line-of-sight) to a sufficient number of anchors cannot be unambiguously localized and is, therefore, said to be in a \emph{blind spot}. In this paper, we analyze the blind spot probability of a typical target by using stochastic geometry to model the randomness in the obstacle and anchor locations. In doing so, we handle correlated anchor blocking induced by the obstacles, unlike previous works that assume independent anchor blocking. We first characterize the regime over which the independent blocking assumption underestimates the blind spot probability of the typical target, which in turn, is characterized as a function of the distribution of the \emph{visible} area, surrounding the target location. Since this distribution is difficult to characterize exactly, we formulate the \emph{nearest two-obstacle approximation}, which is equivalent to considering correlated blocking for only the nearest two obstacles from the target and assuming independent blocking for the remaining obstacles. Based on this, we derive an approximate expression for the blind spot probability, which helps determine the anchor deployment intensity needed for the blind spot probability of a typical target to be at most a threshold, $\mu$.
\end{abstract}


\section{Introduction} \label{sec:intro}
% Applications of indoor low-range high-accuracy localization, (e.g. patient monitoring, smart shopping carts).
% Existing networks are not suitable, GPS does not work for indoors and WiFi and cellular were designed for data traffic. For accurate localization, apart from signal strength, geometry is also important.
% Hence, there is a need for custom localization networks, often using high-bandwidth solutions like mm-wave or UWB.
% Does not have to be active or passive, we can keep the setting very general.
% In all these applications, blocked LoS is very common. In particular, the blocking is correlated. Obstacle close to can block pretty much all the anchors to one side.

%I suggest for this paragraph  to start out with the general ToA localization problem, and only then move to specific examples .

% In GPS-challenged environments (e.g. indoors), it is typically realized by deploying a network of anchors over the region of interest

% Depending on the localization technique used  (e.g., time-of-arrival (ToA), received signal strength (RSSI), angle-of-arrival (AoA) etc.), a target should have line-of-sight (LoS) to at least a minimum number of anchors for unambiguous localization (e.g., for ToA-based localization over a 2D-plane, this number equals three). However, in many applications, the LoS link between a target and an anchor may be blocked by obstacles present in the environment. If a target does not have LoS to the required number of anchors, then a unique estimate of its position cannot be obtained and hence, is said to be in a \emph{blind spot}.

Accurate localization is becoming an increasingly indispensable tool for supporting a variety of indoor and outdoor applications such as wildlife monitoring, location-based advertising \cite{Steiniger_et_al_2006}, search-and-rescue operations \cite{Lo_Xia_etal_2008}, self-driving vehicles \cite{Murrian_et_al_2016}, assisted living \cite{Witrisal_et_al_2016}, remote RFID \cite{DarErr_2008,dardari2010ultrawide,Dec_Gui_Dar_2014} etc. Many of the above applications require high positioning accuracy (between $0.1-1{\rm m}$) in environments where the Global Positioning System (GPS) is traditionally unreliable (e.g., indoors, street canyons etc.). A feasible solution to overcome this challenge is to realize a terrestrial wireless localization network by deploying transceivers, known as anchors, throughout the region of interest. For reasons of cost and energy efficiency, each anchor may be equipped with only a single antenna in some deployments. As a result, directional information such as the angles of arrival/departure cannot be exploited from the signal emanating from a target (e.g., a car or an RFID tag). Under these conditions, a target can be localized over a plane if its distance (also known as \emph{range}) to at least three anchors is known\footnote{Throughout this work, we assume 2D localization for convenience. The extension to the 3D case is straightforward. In particular, the range to at least \emph{four} anchors is required for unambiguous 3D localization.}. The ranges can be estimated from the time-of-arrival (ToA) of a \emph{ranging signal} along the line-of-sight (LoS) path\footnote{This requires the targets and anchors to be synchronized.} and when the available bandwidth is large (e.g., of the order of GHz), ToA-based localization can provide sub-meter accuracy \cite{Gezici_et_al_2005}.
%\begin{figure}
%  \centering
%   \includegraphics[scale=0.35]{toa_example.eps}
%   \caption{ToA based localization: Each distance (range) estimate constrains the target to lie on a circle centered at the corresponding anchor, whose radius equals the range. The intersection of three or more such circles provides an unambiguous solution for the target location in $\nbbR^2$.}
%   \label{fig:trilateration}
%\end{figure}

However, in many of the applications listed above, the LoS link between an anchor-target pair may be blocked by obstacles in the environment. Using vision as an analogy, an anchor is said to be \emph{invisible (visible)} to a target, if the LoS between the anchor and the target is blocked (unblocked). Consequently, for ToA-based localization, a target is said to be in a blind spot if it is visible to fewer than three anchors, since it cannot be localized unambiguously. If the map of the environment is known, then a deterministic, blind spot eliminating placement of anchors can be obtained by solving a variant of the art-gallery problem \cite{Ebra_Scholtz_2005, Gonzalez-Banos:2001:RAA:378583.378674}. However, in many applications, the map of the environment may not be known beforehand; for instance,
\begin{itemize}
 \item In a forest environment where the target(s) are wildlife, the trees could act as obstacles. In this case, it is unreasonable to assume that all obstacle locations are known.
 \item On a road or in a shopping mall, vehicles and humans may respectively act as obstacles intermittently.
\end{itemize}
The above examples represent a diverse range of indoor and outdoor situations, where the obstacles can either be static or dynamic. Additionally, since the obstacles are typically not point objects, the blocking of LoS across multiple links exhibits correlation, in general (e.g., anchors A1 and A2 are blocked to the target by the same obstacle in Fig. \ref{fig:germ_grain}). To the best of our knowledge, the existing literature on the art gallery problem does not address the question of eliminating or minimizing the occurrence of blind spots when the environment map is unknown.

To address this gap, we consider a stochastic geometry based framework where we use random shape theory to model the obstacle locations and shapes and a homogeneous PPP to model the anchor locations\footnote{For a number of commercial applications, the anchors would typically be cellular base stations that also provide other wireless communication services. The PPP is a standard model for base station deployment in wireless communication. Furthermore, for some other applications (e.g., dropping anchors from the air to provide wildlife tracking capability in a forest), a deterministic placement of anchor nodes is inherently impossible and a point process model for the anchor locations is appropriate.}. Apart from capturing the uncertainty in the obstacle locations, random shape theory also enables us to model the correlated blocking phenomenon caused by obstacles of varying sizes and shapes. Ignoring the correlation in LoS blocking events and assuming independent blocking across links instead (as was done in previous papers) can result in the underestimation of the blind spot probability at a given (target) location. For instance, if two anchors, situated close to one another, are each invisible to a target with probability $p$, then the joint blocking probability of the two anchor-target links is also approximately $p$, which exceeds $p^2$, the result obtained by assuming independent blocking. 

Due to the probabilistic nature of the anchor and obstacle locations, it is not possible to completely eliminate blind spots. This motivates the analysis of the \emph{blind spot probability} of a typical target over a localization network, which is a performance measure over an ensemble of environment realizations instead of a particular snapshot. In this paper, we analyze the relationship between the blind spot probability and the statistics of the obstacle sizes and locations and the anchor point process. In doing so, we wish to determine the intensity with which anchors need to be deployed so that the blind spot probability over the entire region is less than a threshold $\mu$.

\subsection{Related Work}  \label{sec:related}
%%%%%%%%%%%%%%%%%%%%%
%For the same system model, the impact of base-station geometry on localization accuracy was studied in \cite{Schlo_Dhill_Buehr_2016}, by analyzing the geometric dilution of precision (GDOP) metric.
The PPP was used to model base station locations while investigating the \emph{hearability} problem for localization in cellular networks \cite{Schlo_Dhill_Buehr_2015, Schlo_Dhill_Buehr_2016_2}, where similar to the visibility analogy, the hearability metric was defined as the number of base stations whose SINR (signal-to-interference-plus-noise ratio) at a target mobile station crossed a particular threshold. However, independent log-normal shadowing was assumed for all links and the blocked LoS senario was not specifically addressed. The Boolean model has been used to analyze the impact of blocking on the performance of urban cellular networks \cite{Bai_Vaze_Heath_2013}, and mm-wave systems \cite{Gapeyenko_etal_2016, Hriba_Valenti_2017, Dong_Kim_2016, Samuylov_etal_2016}. In \cite{Bai_Vaze_Heath_2013, Hriba_Valenti_2017, Dong_Kim_2016}, independent blocking was assumed across different links, while in \cite{Samuylov_etal_2016}, the spatio-temporal correlation between the LoS/NLoS states of two links was investigated. The effect of correlated shadowing on the interference distribution of wireless networks in urban areas was studied in \cite{Bac_Zhang_2015}, using a Manhattan line process to model building locations. In the conference version of this paper \cite{Adi_Har_Mol_2016}, we partially considered the impact of correlated blocking by estimating the blind spot probability at a given (target) location using approximate second-order blocking statistics and in \cite{Adi_Har_Mol_Beh_2017}, the worst-case impact of correlated blocking on the blind spot probability was investigated by considering \emph{infinitely} large obstacles modeled by a line process. 
In general though, to the best of our knowledge, stochastic geometry models for correlated shadowing or blocking in wireless networks is an emerging field.

% New references: \cite{Zhang_Baccelli_Heath_2015}, \cite{Lee_Baccelli_2018}

%In general though, to the best of our knowledge, stochastic geometry models for correlated shadowing or blocking in wireless networks remain a scarcely investigated field. 
 
%%%%%%%%%%%%%%%%%%%%%%%%%%
\subsection{Contributions} \label{sec:contributions}
%%%%%%%%%%%%%%%%%%%%%%%%%%
% Shadow/visible regions
The main contributions of this work are as follows:
 \begin{itemize}
  \item We model the anchor locations using a homogeneous PPP and the obstacle locations and shapes using random shape theory (specifically, a Boolean model). From the perspective of a typical target, the anchors that are within communication range are constrained to lie in a circular region, centered at the target. The obstacles lying within this circle partition it into \emph{visible} and \emph{shadowed} regions, where the anchors lying in the shadowed region are invisible to the target. Under these conditions, we express the blind spot probability at a typical target location in terms of the probability distribution of the visible area (i.e., the area of the visible region surrounding a typical target).
  \item We then show that the blind spot probability under the independent anchor blocking assumption depends only on the \emph{mean} visible area, instead of the entire probability distribution. In addition, we derive the conditions under which the independent blocking assumption underestimates the true blind spot probability.
  \item We then demonstrate that the visible area distribution is difficult to characterize in closed form. As a result, we propose an approximate solution for characterizing the visible area whereby in each environment realization, the visible area is evaluated \emph{exactly} up to the location of the second nearest obstacle and the remaining value beyond that is approximated by its mean. We refer to this as the \emph{nearest two-obstacle approximation} and we show that it is equivalent to considering correlated blocking up to the location of the second nearest obstacle and assuming independent blocking for farther obstacles, where the impact of blocking correlation is relatively minimal. In other words, the nearest two-obstacle approximation engenders a \emph{quasi-independent} blocking assumption.
  \item Using the nearest two-obstacle approximation, we derive a closed-form approximation for the blind spot probability as well as the conditions under which it yields a tighter bound on the true blind spot probability, relative to the independent blocking assumption. As a result, our work provides useful design insights, such as the intensity with which anchors need to be deployed so that the blind spot probability over the entire region is less than a threshold, $\mu$.
 \end{itemize}

\subsection{Notation}
Throughout this work, bold lowercase Latin (e.g., $\nba$) or Greek letters (e.g., $\boldsymbol{\alpha}$) are used to represent vectors. $\nbbR$ denotes the set of real numbers and $\nu_2(.)$ denotes the Lebesgue measure in $\nbbR^2$ (i.e., for a set $\ncalS\subseteq \nbbR^2$, $\nu_2(\ncalS)$ denotes the area of $\ncalS$). The probability of an event $\mathsf{A}$ is denoted by $\nbbP(\mathsf{A})$ and the expectation operator is denoted either by $\nbbE_X[.]$, to explicitly indicate expectation with respect to a random variable, $X$; or by $\nbbE[.]$, when the context is clear. $\bigcup$ and $\bigcap$ denote set union and intersection, respectively, and $\varnothing$ denotes the empty set. A real function $f$, with argument $t$ and parameters given by a vector, $\nba$, is denoted by $f(t;\nba)$. Finally, for a function $f:\nbbR \rightarrow \nbbR$, ${\rm graph}(f) \triangleq \{(x,y)\in \nbbR^2: y = f(x)\}$ and ${\rm epi}(f) \triangleq \{(x,y)\in \nbbR^2: y\geq f(x)\}$ denote its graph and epigraph, respectively \cite{Boyd_cvx}.

\subsection{Organization}
This paper is divided into seven sections. The system model is described in Section~\ref{sec:SysMod}, where the anchor locations are modeled using a homogeneous PPP and the obstacles are represented using line-segments of random lengths and orientations. In Section~\ref{sec:blindspot_analysis}, the blind spot probability at a typical target location is characterized in terms of the distribution of the surrounding visible area. Additionally, the blind spot probability under the independent anchor blocking assumption is also characterized and the conditions under which it underestimates the true blind spot probability are derived. The nearest two-obstacle approximation is introduced in Section \ref{sec:shadow_area} to characterize the visible area in a tractable manner, which is then used to derive an approximate expression for the blind spot probability in Section \ref{sec:tractable_approx}, that takes into account the impact of correlated blocking up to the second nearest obstacle. Numerical results to validate our approximations are presented in Section~\ref{sec:NumResults}. Finally, Section~\ref{sec:concl} concludes the paper.
%%%%%%%%%%%%%%%%%%%%%%
\section{System Model} 
\label{sec:sysmodel}
%%%%%%%%%%%%%%%%%%%%%%

%\label{sec:sysmodel}
%\begin{figure}
%  \centering
%  \includegraphics[scale=0.3]{small_obst}
%  \caption{A localization network consisting of anchors ($\bigtriangleup$) and obstacles ($\diagup$) surrounding a target. The distributed obstacles give rise to correlated blocking. The shadow regions can be viewed as a germ-grain model, where the germs are the obstacle mid-points and the grains are the shaded regions.}
%  \label{fig:germ_grain}
%\end{figure}

%A. Describe the spatial model in detail (for targets, obstacles, etc.)
\begin{figure}
  \centering
 \begin{subfigure}{0.59\textwidth}
  \includegraphics[scale=0.4]{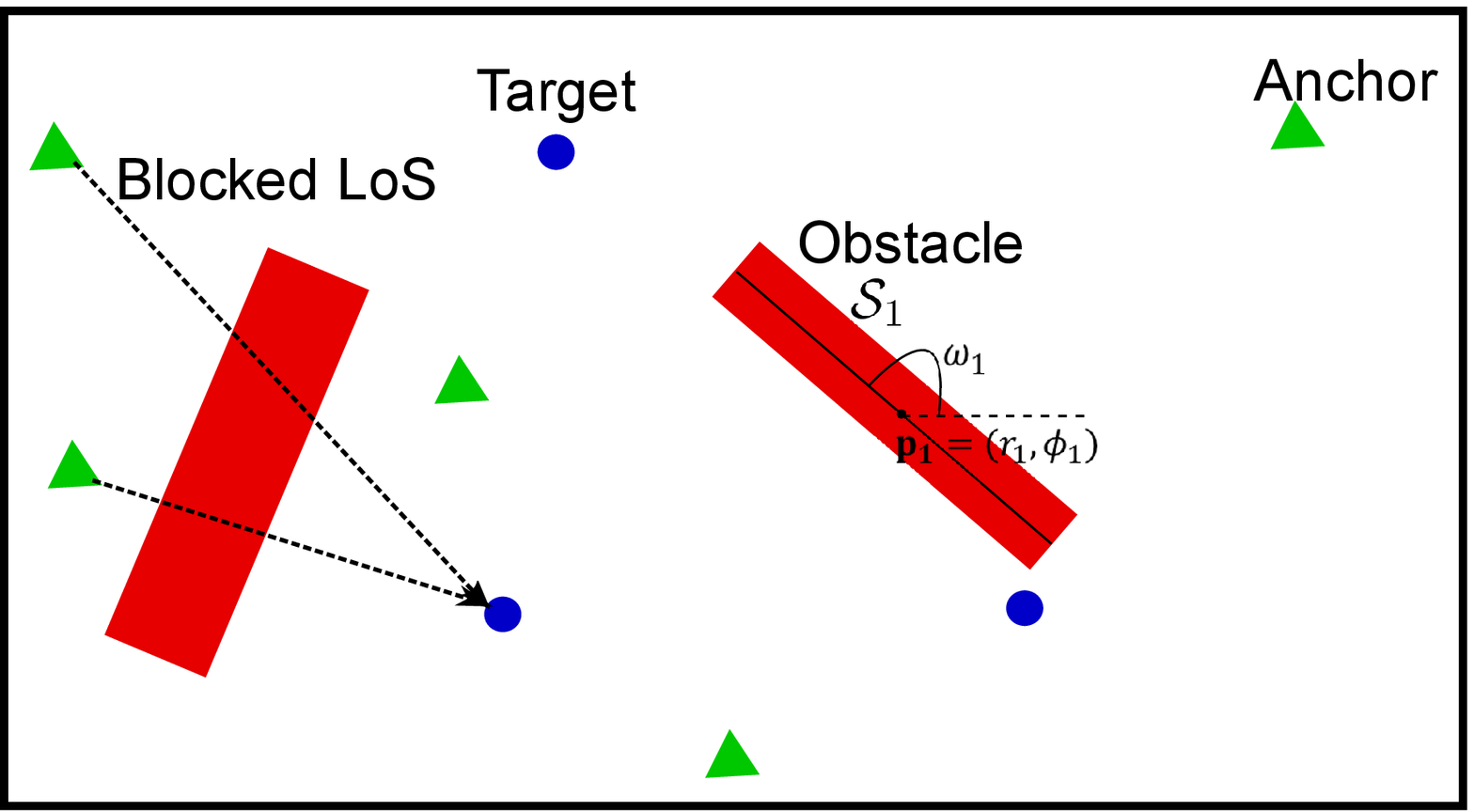}
  \caption{Example of a localization scenario consisting of anchors, targets and obstacles.}
  \label{fig:big_pic}
 \end{subfigure}
 \hfill
  \begin{subfigure}{0.39\textwidth}
  \centering 
  \includegraphics[scale=0.25]{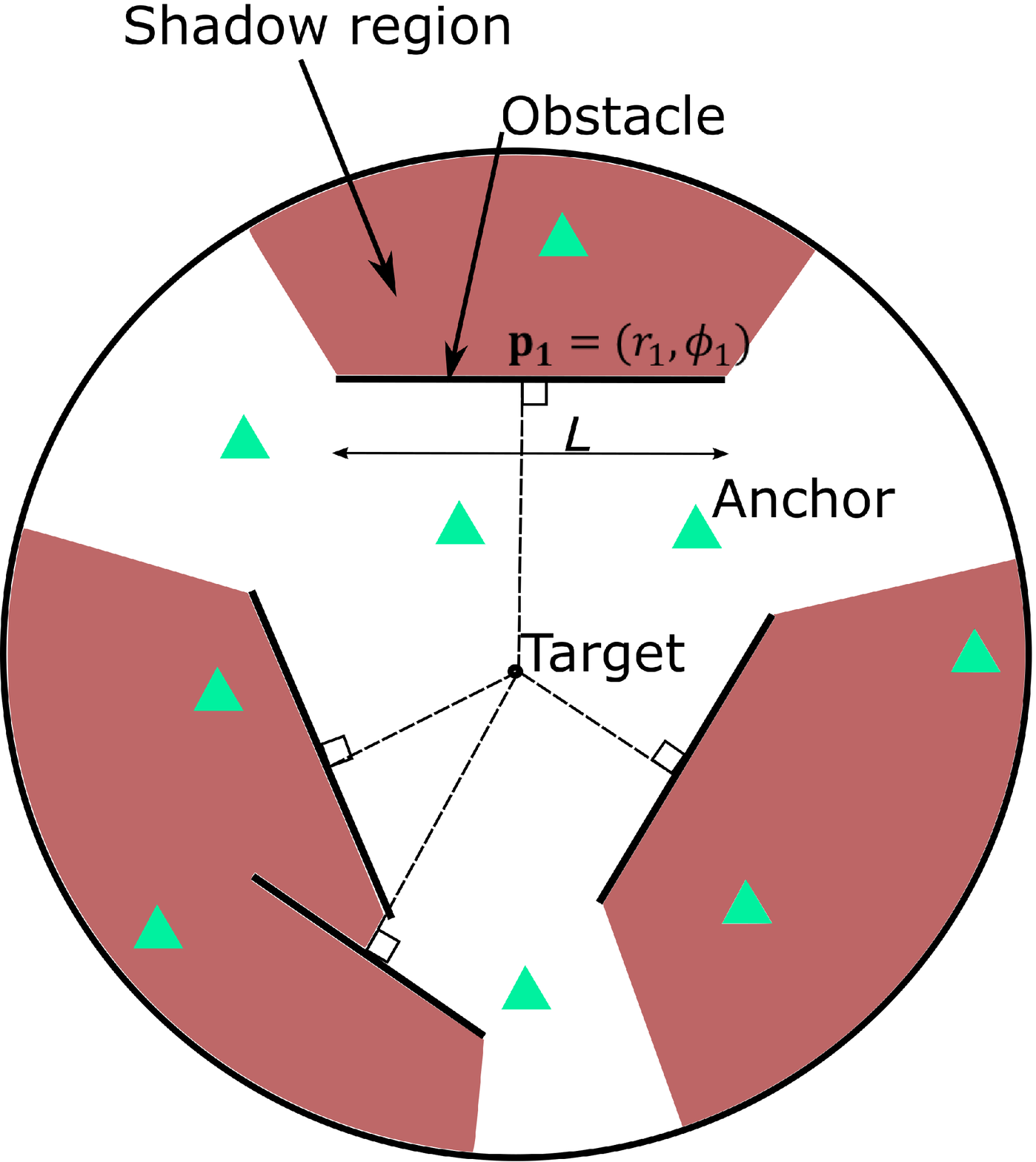}
  \caption{Visible region around a typical target, for the line segment obstacle model where all the obstacles have length $L$ and face the target $(\omega_i=\phi_i+\pi/2)$.}
  \label{fig:germ_grain}
 \end{subfigure} 
 \caption{Illustration of the stochastic geometry based system model.}
 \label{fig:illus}
\end{figure}

Consider an environment in $\mathbb{R}^2$ consisting of point targets and distributed obstacles. Intuitively, the $i$-th obstacle can be parametrized by the tuple $(\nbp_i,\ncalS_i,\omega_i)$, where $\ncalS_i\subseteq \mathbb{R}^2$ denotes the `shape' of the obstacle (e.g., a rectangle), $\nbp_i=(r_i,\phi_i)\in \mathbb{R}^2$ its `location' in polar coordinates $(\phi_i \in [0,2\pi))$ (e.g., the geometric center), and $\omega_i \in [0,2\pi)$ its `orientation' with respect to the positive $x$-axis, as shown in Fig.~\ref{fig:big_pic}. The collection of obstacles, $\bigcup\limits_i (\nbp_i,\ncalS_i,\omega_i)$, forms a germ-grain model if the following conditions are satisfied \cite{Sto_et_al_2013}:
\begin{itemize}
 \item[(i)] The set of points $\{\nbp_i\}$, known as germs, form a point process in $\mathbb{R}^2$.
 \item[(ii)] The set $\{(\ncalS_i,\omega_i)\}$, known as grains are drawn from a family of closed sets $\nbbS\times\Omega$.
\end{itemize}
The obstacles are assumed to be opaque to radio waves; therefore, the obstacle \emph{thickness} does not influence the existence of LoS and hence, it is sufficient to let $\nbbS$ be the set of line-segments of length at most $L$, where $L$ is the maximum obstacle length (i.e., $\nbbS \triangleq [0,L]$). Without loss of generality, the germs can be chosen to be the mid-points of the line-segments\footnote{In general, the germs need not be the geometric centers of their corresponding grains.}. Thus, $\Omega \triangleq [0,\pi)$ is sufficient to encompass all obstacle orientations (e.g., Fig.~\ref{fig:big_pic}). We assume the germs to be distributed according to a homogeneous PPP with intensity $\lambda_0$. The obstacle lengths and orientations are modeled as samples drawn from a joint distribution, supported on $\nbbS \times \Omega$, whose probability density function (pdf) is denoted by $f_{\mathsf{L},\mathsf{W}}(\cdot,\cdot)$, where $\mathsf{L}$ and $\mathsf{W}$ denote the random variables representing the obstacle length and orientation, respectively. 

%B. Explain why we confine to a circle (limited range, etc.)
A localization network comprising of single-antenna anchors is deployed over $\mathbb{R}^2$ and we assume the anchor locations to also form a homogeneous PPP, with intensity $\lambda$, independent of the obstacle germ process. ToA-based localization is assumed throughout and we assume that the targets transmit a ranging signal omnidirectionally\footnote{We assume that the targets employ a medium access control protocol to coordinate their transmissions in order to avoid interference.}, which is received at the anchors and used for ToA/range estimation and subsequent localization.

\begin{figure}
 \centering
  \includegraphics[scale=0.4]{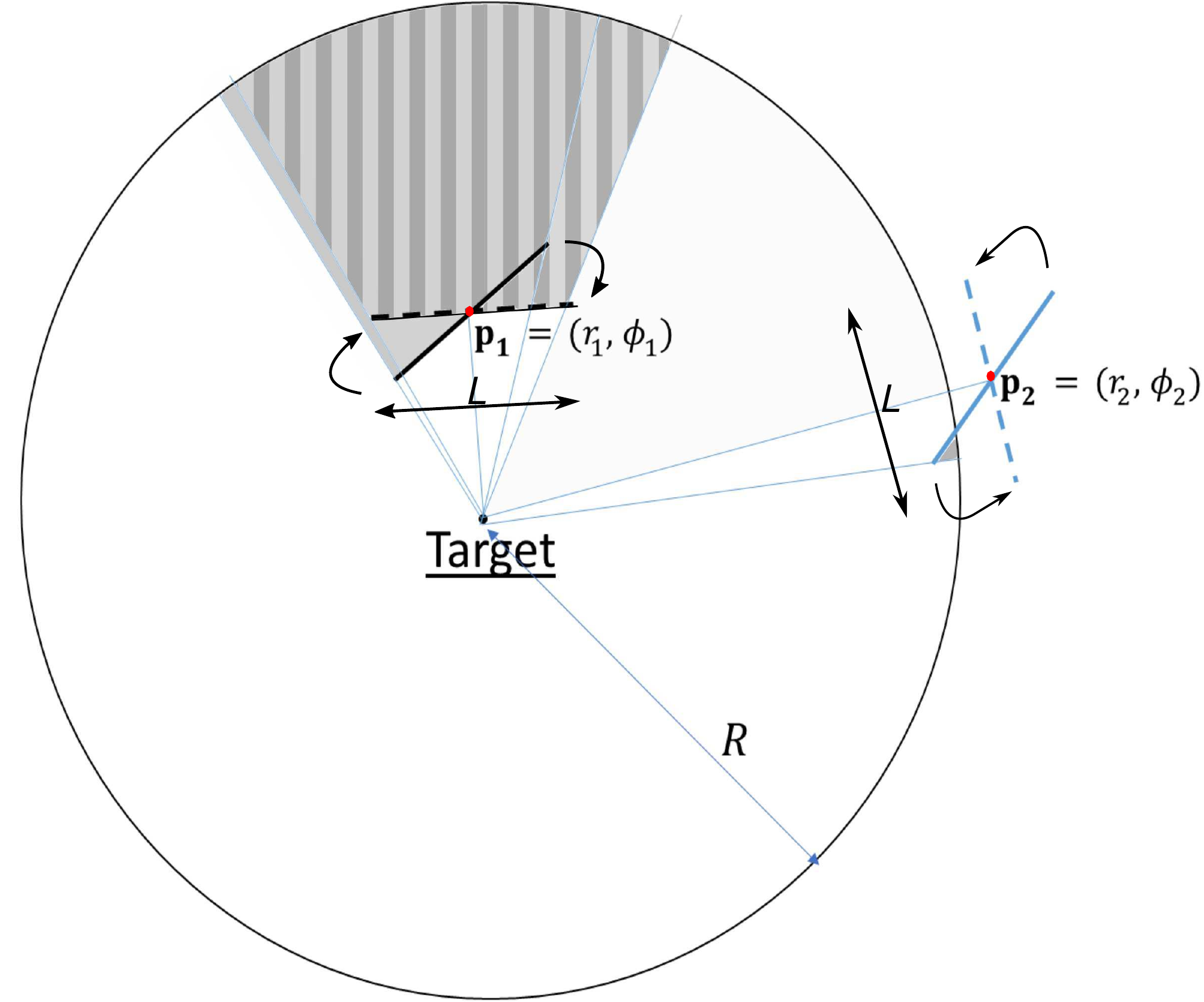}
  \caption{Illustration of the quasi worst-case obstacle orientation, where all obstacles are assumed to \emph{face} the typical target (illustrated using dotted lines) and have maximum length, $L$. This maximizes the shadowed area due to obstacles whose mid-points are within $\ncalD_\nbo(R)$ (e.g., $\nbp_1$ above. The shadow regions due to the original and `rotated' orientations are represented using the plain and striped grey regions, respectively.), while neglecting the shadowed region induced by obstacles whose mid-points lie outside $\ncalD_\nbo(R)$ (e.g., $\nbp_2$ above).}
  \label{fig:shadow}
\end{figure}

Due to the stationarity of the PPP, it can be assumed without loss of generality that a target is situated at the origin, $\nbo$, which we refer to as the \emph{typical} target. A transmit power constraint further restricts our attention to a disc of radius $R$, centered around $\nbo$ and denoted by $\ncalD_\nbo(R)$, in which anchors must lie for the target to be localized. From the target's perspective, each obstacle induces a shadow region, which is the set of points that it renders invisible to the target, as illustrated in Fig.~\ref{fig:germ_grain}. Consequently, the anchors that lie in a shadow region are invisible to the target. The shadow regions form a germ-grain model (Fig. \ref{fig:germ_grain}), where the area of a grain depends on how far its germ (i.e., the corresponding obstacle mid-point) is from $\nbo$. Since $f_{\mathsf{L},\mathsf{W}}(\cdot,\cdot)$ is usually unknown, we assume all obstacle lengths are equal to $L$ and $\omega_i=\phi_i+\pi/2$ (see Fig. \ref{fig:germ_grain}). If $r_i\leq R$ (e.g., the obstacle with mid-point $\nbp_1$ in Fig. \ref{fig:shadow}), then such a rotation of the $i$-th obstacle to \emph{face} the (typical) target maximizes the area of its shadow region; on the other hand, if $r_i>R$ (e.g., the obstacle with mid-point $\nbp_2$ in Fig. \ref{fig:shadow}), this rotation eliminates any shadow region due to the $i$-th obstacle, thereby ignoring the blocking caused by it. As a result, this assumption corresponds to a \emph{quasi worst-case} orientation for the obstacles that emphasizes the (greater) influence of nearer obstacles on correlated anchor blocking and subsequently, the blind spot probability. 
%D. Explain the metric of interest and why that is useful.
Thus, the obstacles whose mid-points lie within $\ncalD_\nbo(R)$ partition it into \emph{shadowed} and \emph{visible} regions, where for ToA-based localization, the target can be localized if it there are at least three anchors in the visible region. Consequently, the target is said to be in a \emph{blind spot} if this condition is not satisfied. As blind spots are undesirable, the blind spot probability of the typical target is an important metric from a network design perspective. In the following section, we develop the relationship between the blind spot probability at a typical target location and its surrounding visible area distribution, which is a function of the obstacle intensity ($\lambda_0$) and size ($L$).
\begin{remark}
The stationarity of the anchor and obstacle germ PPPs ensure that the statistics of the visible region surrounding any target location is the same. Hence, even if multiple targets are present (e.g., Fig.~\ref{fig:big_pic}), it is sufficient to analyze the single target case in order to bound the blind spot probability at \emph{all} target locations. This helps define the notion of a \emph{typical} target at the origin. 
\end{remark}
%%%%%%%%%%%%%%%%%%%%%%%%%%%%%%%%%
\section{Analysis of blind spot Probability}
 \label{sec:blindspot_analysis}
%%%%%%%%%%%%%%%%%%%%%%%%%%%%%%%%%
For the parameter vector $\nbz=[\lambda_0~ L ~ R]$, we define the \emph{visibility} random variable, denoted by $V(\nbp;\nbz)$ for $\nbp= (r,\phi)\in \ncalD_\nbo(R)$, in the following manner:  
\begin{align}
 \label{eq:visible}
V(\nbp;\nbz)&=\begin{cases}
			& 1, \mbox{ if $\nbp$ is visible to $\nbo$} \\
 			& 0, \mbox{ else.}
 		   \end{cases}
\end{align}
Let $\ncalV(\nbz)=\{\nbp \in \ncalD_\nbo(R): V(\nbp;\nbz)=1 \}$ denote the visible region around the target and let $A_v(\nbz)=\nu_2(\ncalV(\nbz))$ denote its area, which we refer to as the \emph{visible area}. The typical target is in a blind spot if and only if there are fewer than three anchors in $\ncalV(\nbz)$. Thus, the blind spot probability, conditioned on the random variable $A_v(\nbz)$ and denoted by $g(A_v(\nbz);\lambda)$, with parameter $\lambda$, has the following expression:
\begin{align}
\label{eq:condprob_BS}
 g(A_v(\nbz);\lambda) &\triangleq \nbbP(\mbox{blind spot }|~A_v(\nbz)) \notag \\
 &= \displaystyle\sum\limits_{k=0}^2 \nbbP(k \mbox{ anchors present in the visible region of area $A_v(\nbz)$}) \\
 &= e^{-\lambda A_v(\nbz)}\left(1+\lambda A_v(\nbz) +\frac{(\lambda A_v(\nbz))^2}{2}\right).
\end{align}
\begin{remark}
The definition of a blind spot can be generalized to the absence of at least $k_{\rm v}$ visible anchors in $\ncalV(\nbz)$, due to which the summation limits in (\ref{eq:condprob_BS}) would run from 0 to $k_{\rm v}-1$. This is useful to analyze the blind spot probability for other localization techniques such as time-difference of arrival (TDoA) based localization for which $k_{\rm v}=4$. 
\end{remark}

The unconditional blind spot probability, $b(\lambda,\nbz)$, is then obtained by averaging over the distribution of $A_v(\nbz)$ as given below,
\begin{align}
  \label{eq:prob_BS}
  b(\lambda,\nbz) = \displaystyle\int\limits_0^{\pi R^2} g(t;\lambda) f_{A_v(\nbz)}(t) {\rm d}t,
\end{align}
where $f_{A_v(\nbz)}(.)$ is the pdf of $A_v(\nbz)$, which fully captures the statistics of correlated anchor blocking due to obstacle size $L$ and intensity $\lambda_0$.

The visible anchors can be interpreted as a point process derived by sampling from the underlying anchor PPP, where an anchor at point $\nbp$ is selected with a probability equal to $\nbbP(V(\nbp;\nbz)=1)$. Furthermore, the sampling process is also correlated across anchor locations due to correlated blocking (i.e., the probability that an anchor at $\nbp$ is selected also depends on the selection of other anchors in $\ncalD_\nbo(R)$). However, if we ignore this correlation and assume that each anchor is sampled independently of the other anchors, we obtain the well-known \emph{independent blocking} assumption, for which the resulting blind spot probability is given by the following lemma:
\begin{lemma} \label{lem:pbs_indep}
The blind spot probability under the independent anchor blocking assumption, denoted by $b^{\rm ind}(\lambda,\nbz)$, is given by:
\begin{align}
 \label{eq:probBS_indep}
 b^{\rm ind}(\lambda,\nbz)&=e^{-\lambda \nbbE[A_v(\nbz)]}\left(1+\lambda \mathbb{E}[A_v(\nbz)]+\frac{(\lambda \mathbb{E}[A_v(\nbz)])^2}{2}\right) = g(\mathbb{E}[A_v(\nbz)];\lambda).
\end{align}
\end{lemma}
\begin{IEEEproof}
See Appendix~\ref{app:pbs_indep}.
\end{IEEEproof}
%\begin{cor}\label{cor:LLN}
%$f_{A_v(\nbz)}(t)=\delta(t - \nbbE [A_v])$ for independent anchor blocking, where $\delta(.)$ denotes the Dirac-delta function.
%\end{cor}
%\begin{IEEEproof}
% See Appendix \ref{app:LLN}.
%\end{IEEEproof}

From Lemma \ref{lem:pbs_indep}, it can be seen that the mean visible area, $\nbbE[A_v(\nbz)]$, completely characterizes the blind spot probability if independent anchor blocking is assumed. For the system model from Section \ref{sec:sysmodel}, $\nbbE[A_v(\nbz)]$ is given by the following lemma:
\begin{lemma} \label{lem:Ash_firstmom}
For a parameter vector $\nbz$, the average visible area, $\nbbE[A_v(\nbz)]$, over $\ncalD_\nbo(R)$ is given by:
 \begin{align}
  \nbbE [A_v(\nbz)]&=2\pi \displaystyle\int\limits_0^{R} \exp(-\lambda_0 \nu_2(\ncalS_V(\nbp;\nbz))) ~ r {\rm d}r {\rm d}\phi, \\
  \label{eq:Sv_first}
  \mbox{where } \ncalS_V(\nbp;\nbz)&= \{(\rho,\beta) \in \nbbR^2: 0 \leq \rho \tan |\beta-\phi| \leq L/2, 0 \leq \rho \sec |\beta-\phi| \leq r\}\\
  \mbox{and }\nu_2(\ncalS_V(\nbp;\nbz))&= 2\displaystyle\int\limits_0^r \rho \min\left(\arctan\left(\frac{L}{2\rho}\right), \arccos\left(\frac{\rho}{r}\right)\right)~ {\rm d}\rho.
 \end{align}  
\end{lemma}
\begin{IEEEproof}
 See Appendix~\ref{app:Ash_firstmom}.
\end{IEEEproof}

The relationship between $b(\lambda,\nbz)$ and $b^{\rm ind}(\lambda,\nbz)$ is given by the following theorem:
\begin{theorem} \label{thm:jensens}
$b^{\rm ind}(\lambda,\nbz) \leq b(\lambda,\nbz)$ over $\{(\lambda,\nbz):\lambda\mathbb{E}[A_v(\nbz)]\geq 3.3836 \}$.
\end{theorem}
\begin{IEEEproof}
As a twice-differentiable function of $t$, we have
 \begin{align}
  \label{eq:fder_g}
\frac{{\rm d}}{{\rm d} t}g(t;\lambda) &= -(\lambda^3/2) t^2 e^{-\lambda t} \\
  \label{eq:sder_g}
 \frac{{\rm d}^2 }{{\rm d}^2 t}g(t;\lambda) &= (\lambda^3/2)t e^{-\lambda t}(\lambda t - 2).
 \end{align}
From (\ref{eq:sder_g}), the second derivative of $g(t;\lambda)$ is non-negative when $t \geq 2/\lambda$. Hence, $g(t;\lambda)$ is a convex function in $t$ over this regime \cite{Boyd_cvx}. Let $t_0$ denote the solution to the following equation:
\begin{align}
 \label{eq:t0_eqn}
 1&=g(0;\lambda)=g(t_0;\lambda)-t_0  \frac{{\rm d} }{{\rm d} t}g(t;\lambda)\bigg|_{t=t_0} \\
 \label{eq:t0_eqn_expanded}
 \implies~ 1 &= e^{-\lambda t_0}\left[\frac{(\lambda t_0)^3}{2} +\frac{(\lambda t_0)^2}{2} + \lambda t_0 + 1 \right].  
\end{align}
Eqn. (\ref{eq:t0_eqn_expanded}) is a mixed polynomial-exponential equation in $\lambda t_0$ and solving for $\lambda t_0$ numerically, we obtain (up to four digits of precision),
\begin{align}
 t_0 &= \frac{3.3836}{\lambda}.
\end{align}
\begin{figure}
 \centering
 \includegraphics[scale=0.7]{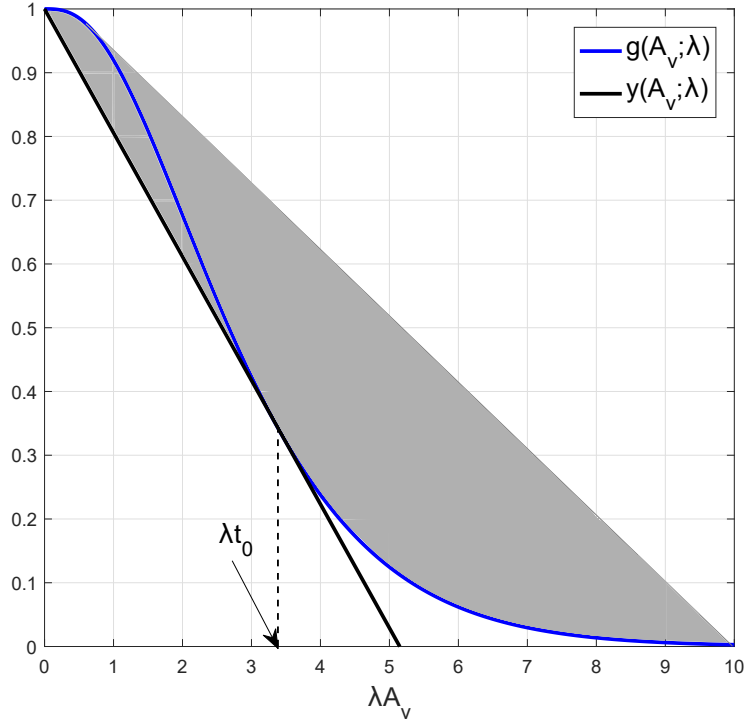}
 \caption{For $\lambda \nbbE[A_v(\nbz)]\geq \lambda t_0 = 3.3836$, the set of points $\{(\nbbE[A_v(\nbz)], b(\lambda,\nbz))\}$ (i.e., the grey shaded region) lies above the set $\{(\nbbE[A_v(\nbz)], b^{\rm ind}(\lambda,\nbz))\}$, shown by the blue curve.}
 \label{fig:Jensen_proof}
\end{figure}
Geometrically, $t_0$ determines the $x$-coordinate of the point at which the line $y(t;\lambda)\subseteq \nbbR^2$, passing through $(0,g(0;\lambda))$, is tangential to ${\rm epi}(g(.;\lambda))$, as shown in Fig. \ref{fig:Jensen_proof}. The equation of $y(t;\lambda)$ is as follows:
\begin{align}
 y(t;\lambda)&=g(t_0;\lambda) + (t-t_0)\frac{{\rm d} }{{\rm d} t}g(t;\lambda)\bigg|_{t=t_0} , ~ t \geq 0.
\end{align}
Let
\begin{align}
 \label{eq:gcon}
 g_{\rm con}(t;\lambda) &= \begin{cases}
 										   & g(t;\lambda) , \hspace{5mm} t > t_0 \\
 										   & y(t;\lambda) , \hspace{5mm} t \leq t_0.
 										   \end{cases}
\end{align}
For $0 \leq t \leq t_0$, the supporting hyperplane at each point, $(t,g_{\rm con}(t;\lambda))$, on the boundary of ${\rm epi}(g_{\rm con}(.;\lambda))$ is $y(.;\lambda)$. Similarly, there also exists a supporting hyperplane at each boundary point, $(t,g_{\rm con}(t;\lambda))$, of ${\rm epi}(g_{\rm con}(.;\lambda))$ for $t>t_0$, since $g_{\rm con}(;\lambda) \equiv g(;\lambda)$, a convex function in its argument over this interval \cite{Boyd_cvx}. Thus, $g_{\rm con}(t;\lambda)$ is a convex function in $t$ for $t\geq 0$. Consequently, if $\nbbE[A_v(\nbz)] > t_0$, then
\begin{align}
\label{eq:pbs_ind}
 b^{\rm ind}(\lambda,\nbz)&= g(\nbbE[A_v(\nbz)];\lambda) 
 \overset{a}= g_{\rm con}(\nbbE[A_v(\nbz)];\lambda)  \overset{b}\leq \nbbE[g_{\rm con}(A_v(\nbz);\lambda)]\overset{c}\leq \nbbE[g(A_v(\nbz);\lambda)]\overset{d}= b(\lambda,\nbz),
\end{align}
%\begin{align}
%\label{eq:pbs_ind}
% b^{\rm ind}(\lambda,\nbz)&= g(\nbbE[A_v(\nbz)];\lambda) \\
%\label{eq:a} 
% &\overset{a}= g_{\rm con}(\nbbE[A_v(\nbz)];\lambda) \\
%\label{eq:b} 
% &\overset{b}\leq \nbbE[g_{\rm con}(A_v(\nbz);\lambda)] \\
%\label{eq:c} 
% &\overset{c}\leq \nbbE[g(A_v(\nbz);\lambda)] \\
%\label{eq:d} 
% &\overset{d}= b(\lambda,\nbz),
%\end{align}
where $(a)$ follows from (\ref{eq:gcon}), $(b)$ from Jensen's inequality, $(c)$ from the fact that $g_{\rm con}(t;\lambda) \leq g(t;\lambda)$ for all $t\geq 0$, and $(d)$ from the definition of $b(\lambda,\nbz)$ in (\ref{eq:prob_BS}). 
\end{IEEEproof}

\begin{remark}
A geometric interpretation of Theorem \ref{thm:jensens} is seen in Fig. \ref{fig:Jensen_proof}, where, as a result of $(d)$ in \ref{eq:pbs_ind}, the feasible values for the ordered pair $(\nbbE[A_v(\nbz)], b(\lambda,\nbz))$ is given by the convex hull of ${\rm graph}(g(.;\lambda))$. On the other hand, the feasible values for $(\nbbE[A_v(\nbz)], b^{\rm ind}(\lambda,\nbz))$ is ${\rm graph}(g(.;\lambda))$, which forms the lower boundary of its convex hull when $\lambda\nbbE[A_v(\nbz)]\geq \lambda t_0 = 3.3836$. It is important to note that Theorem \ref{thm:jensens} represents a sufficient, but not necessary, condition as the proof is a consequence of the convexity properties of $g(\cdot;\lambda)$ that do not depend on $f_{A_v(\nbz)}(.)$. Thus, the inequality $b^{\rm ind}(\lambda,\nbz) \leq b(\lambda,\nbz)$ may still hold over a set $\ncalZ \supseteq \{(\lambda,\nbz):\lambda\mathbb{E}[A_v(\nbz)]\geq 3.3836 \}$ for some choice(s) of $f_{A_v(\nbz)}(.)$. \end{remark}
%\footnote{In the preliminary version of this paper \cite{Adi_Har_Mol_2016}, it was incorrectly stated that $b^{\rm ind}(\lambda,\nbz) \leq b(\lambda,\nbz)$ over $\{(\lambda,\nbz):\lambda\mathbb{E}[A_v(\nbz)]\geq 2 \}$.} 

From a design perspective, it is desirable to have at least three unblocked anchors, on average (i.e., $\lambda \mathbb{E}[A_v(\nbz)]\geq 3$). Hence, from Theorem~\ref{thm:jensens}, it is clear that the
independent blocking assumption underestimates the true blind spot probability for most practical scenarios and that correlated blocking should be taken into account while designing a localization network that meets a desired blind spot probability threshold. From (\ref{eq:condprob_BS})-(\ref{eq:prob_BS}), it is evident that the distribution of the visible area plays a critical role in determining the blind spot probability of the typical target, for a given anchor intensity $\lambda$. In the next section, we attempt to characterize this distribution.

%In the next section, we show how the nearest two-obstacle approximation, proposed in (\ref{eq:Av_assump}) can be used to better approximate the blind spot probability.

\section{Characterizing the visible Area} 
\label{sec:shadow_area}
The visible area around the typical target depends on the number of obstacles as well as their locations. To capture this dependence, we define the following:

%A. Moments of the shadowed area
\begin{definition}
 Let $\ncalV(\nbp^{(k)};\nbz)$ denote a realization of $\ncalV(\nbz)$ when $k(>0)$ obstacle(s) are present, with the obstacle locations determined by $\nbp^{(k)}= [\nbr^{(k)} ~ \nbphi^{(k)}]$, where $\nbr^{(k)}=[r_1 ~ \cdots ~ r_k]$ $(r_i\leq r_j, i<j)$, $\nbphi^{(k)}=[\phi_1 ~ \cdots ~ \phi_k]$, and the $i$-th nearest obstacle mid-point is located at 
$(r_i,\phi_i)$, $(i=1,\cdots,k)$. The special case when $k=0$ is denoted by $\ncalV(\varnothing;\nbz)$ and is equal to $\ncalD_\nbo(R)$.
\end{definition}  

\begin{definition}
 Let $A_v^{(k)}(\nbp^{(k)};\nbz)$ denote the visible area corresponding to $\ncalV(\nbp^{(k)};\nbz)$ (i.e., $A_v^{(k)}(\nbp^{(k)};\nbz) \triangleq \nu_2(\ncalV(\nbp^{(k)};\nbz))$). In particular, $A_v^{(k)}(\nbp^{(k)};\nbz)$ is a realization of the random variable $A_v(\nbz)$, conditioned on the presence of $k$ obstacles whose locations are given by $\nbp^{(k)}$.
\end{definition}

\begin{figure}
 \centering
 \begin{subfigure}{0.45\textwidth}
  \centering
  \includegraphics[scale=0.2]{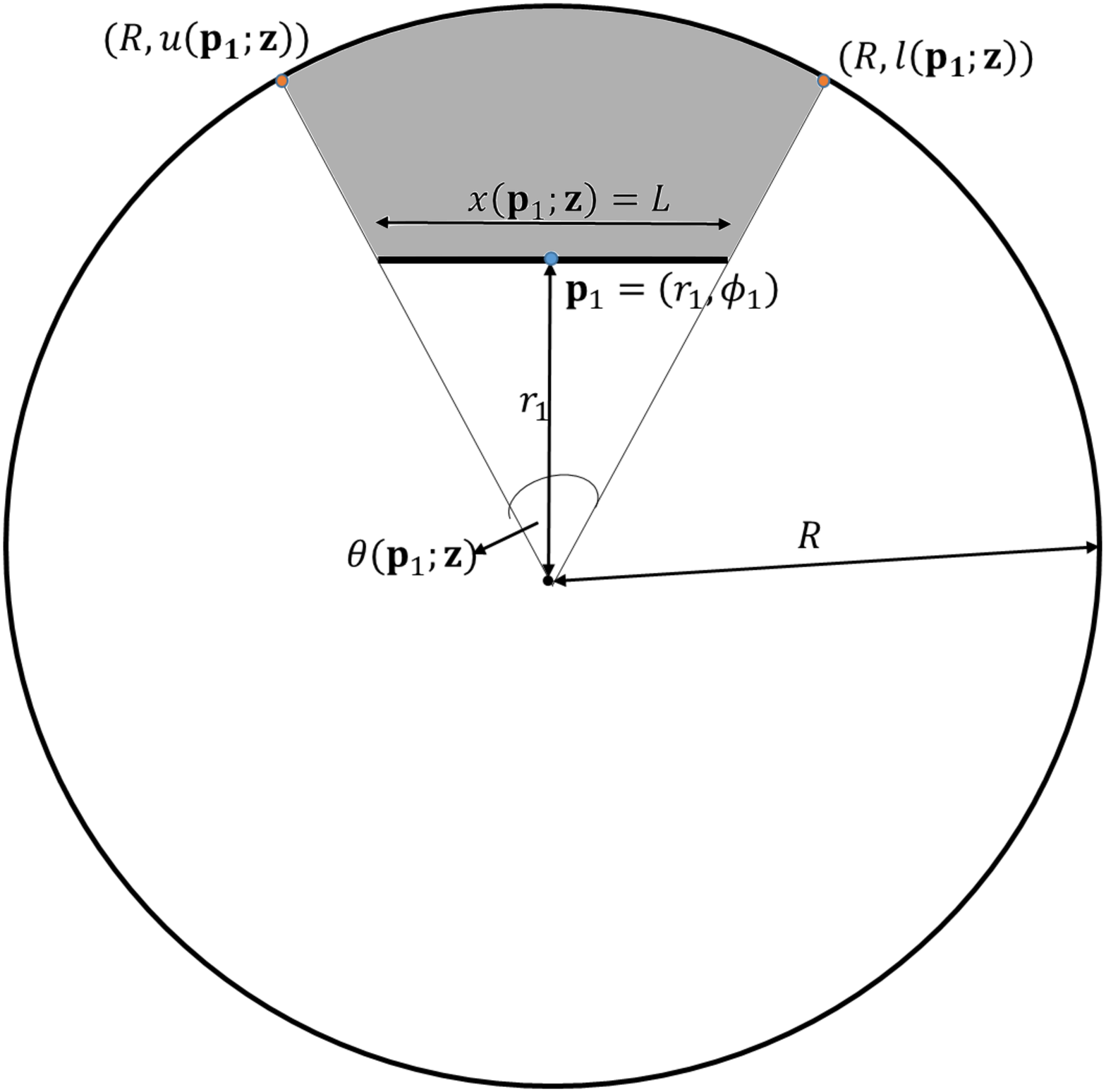}
  \caption{Entire obstacle causes blocking}
 \end{subfigure}
~ \begin{subfigure}{0.45\textwidth}
  \centering
   \includegraphics[scale=0.2]{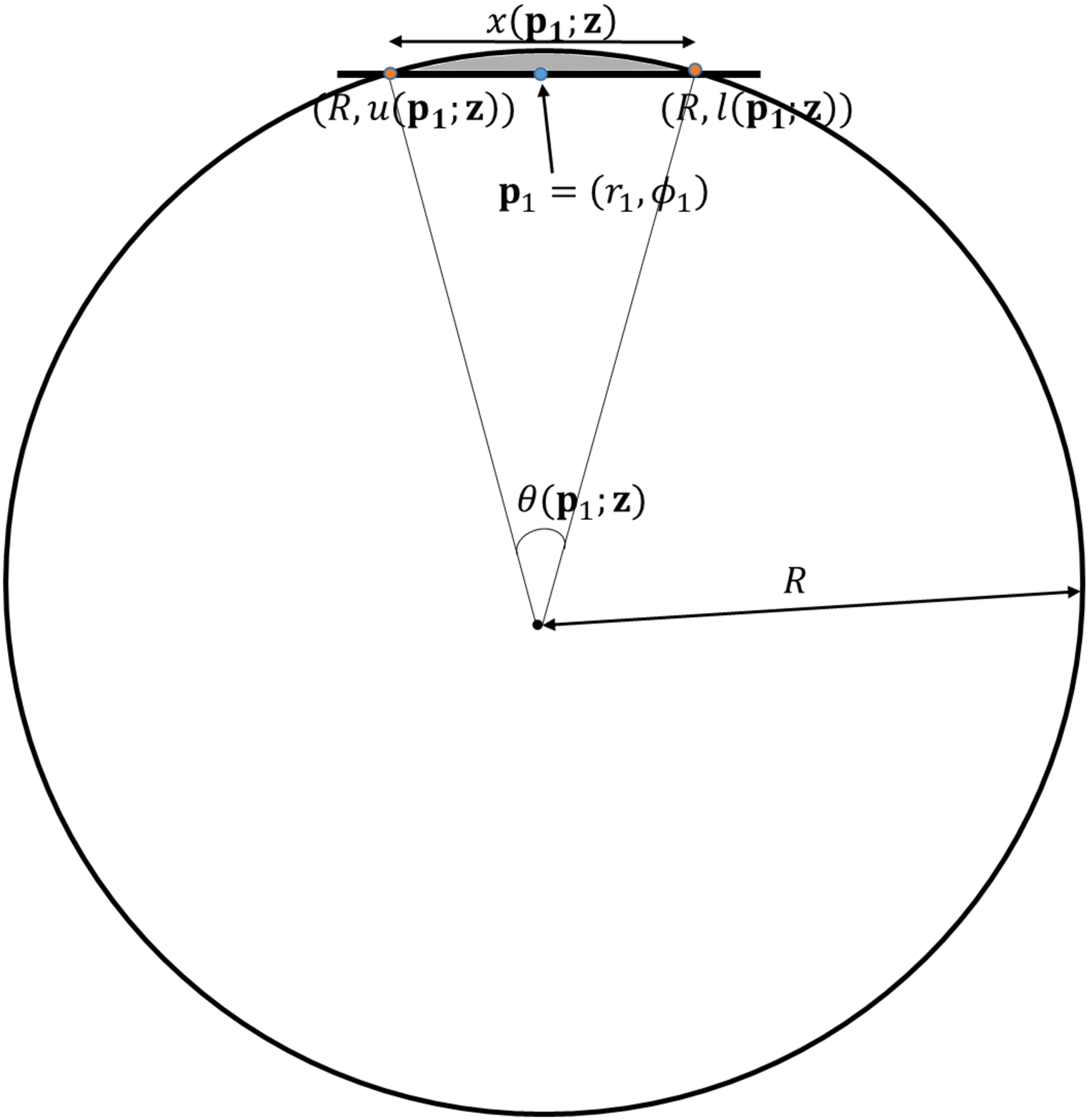}
  \caption{Only a part of the obstacle causes blocking}
 \end{subfigure}
 \caption{Shadowed area (shaded gray) due to a single obstacle.}
 \label{fig:Av_1obstacle}
\end{figure}

For $k<2$, $A_v^{(k)}(\nbp^{(k)};\nbz)$ is easy to characterize,
\begin{align}
 A_{v}^{(0)}(\varnothing;\nbz)&= \pi R^2 \\
 \label{eq:Av_1obstacle}
 A_{v}^{(1)}(\nbp_1;\nbz)&=\pi R^2 -\underbrace{\left( \frac{\theta(\nbp_1;\nbz)}{2}R^2 -\frac{1}{2}r_1x(\nbp_1;\nbz) \right)}_{\text{Shadowed area}}, \\
 \label{eq:theta_exp}
 \mbox{where}~ \theta(\nbp_1;\nbz)&=\begin{cases}
 								2\arctan \left(\frac{L}{2r_1}\right)~,~ 0 \leq r_1 \leq \sqrt{R^2-(L/2)^2} \\
 								2\arccos\left( \frac{r_1}{R} \right)~,~ \sqrt{R^2-(L/2)^2} \leq r_1 \leq R
 							  \end{cases} \\
 				x(\nbp_1;\nbz) &= \begin{cases}
 								L ~,~ 0 \leq r_1 \leq \sqrt{R^2-(L/2)^2} \\
 								2\sqrt{R^2-r_1^2} ~,~ \sqrt{R^2-(L/2)^2} \leq r_1 \leq R.
 							  \end{cases}			  
\end{align}
In particular, the term in parenthesis in (\ref{eq:Av_1obstacle}) denotes the shadowed area (Fig.~\ref{fig:Av_1obstacle}).

For $k\geq 2$, overlaps may occur between the shadow regions corresponding to different obstacles (see Fig.~\ref{fig:germ_grain}). In order to accurately determine $A_{v}^{(k)}(\nbp^{(k)};\nbz)$, the areas of all overlapping shadowed regions should be counted exactly once. We first attempt to characterize the shadow region overlap corresponding to the nearest two obstacles, for which we define the following:

\begin{definition} \label{def:Ash}
 Let $\ncalA_{\rm sh}(\nbp;\nbz)\subseteq \ncalD_\nbo(R)$ denote the shadow region induced by an obstacle whose mid-point is at $\nbp$ (e.g., Fig.~\ref{fig:Av_1obstacle}). The azimuthal end-points of $\ncalA_{\rm sh}(\nbp;\nbz)$, denoted by $l(\nbp;\nbz)$ and $u(\nbp;\nbz)$, are given by the following expressions:
 \begin{align}
  \label{eq:l1}
	 l(\nbp;\nbz)&=\left(\phi-\frac{\theta(\nbp;\nbz)}{2}\right) \mod 2\pi \\
  \label{eq:u1} 
	 u(\nbp;\nbz)&=\left(\phi+\frac{\theta(\nbp;\nbz)}{2}\right) \mod 2\pi, 
 \end{align}
where $\theta(\nbp;\nbz)$ is given by (\ref{eq:theta_exp}). Thus, the azimuthal span of $\ncalA_{\rm sh}(\nbp;\nbz)$, denoted by the interval $\ncalI(\nbp;\nbz) \subseteq [0,2\pi)$, has the following expression:
\begin{align}
\ncalI(\nbp;\nbz) &= [\min(l(\nbp;\nbz),u(\nbp;\nbz)),\max(l(\nbp;\nbz),u(\nbp;\nbz))].
\end{align} 
\end{definition}

\begin{figure}
 \centering
 \includegraphics[scale=0.35]{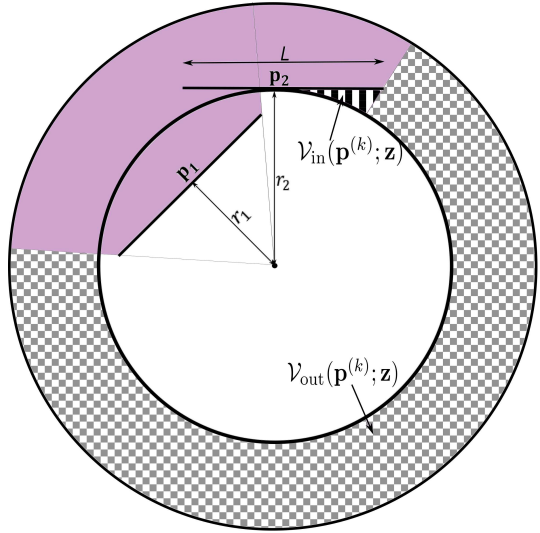}
 \caption{The shaded region denotes the area shadowed by the nearest two obstacles. The additional shadow region induced by the third nearest obstacle onwards must intersect the part-annular checkered region.}
 \label{fig:quasi_setup}
\end{figure}
A typical overlap between a pair of shadow regions $\ncalA_{\rm sh}(\nbp_1;\nbz)$ and $\ncalA_{\rm sh}(\nbp_2;\nbz)$ is illustrated in Fig. \ref{fig:quasi_setup} and the extent of overlap can be characterized by the following lemma.

\begin{lemma} \label{lem:alpha}
 Let $\alpha(\nbp^{(2)};\nbz) \in [0,1]$ denote the fraction of $\ncalA_{\rm sh}(\nbp_2;\nbz)$ that overlaps with $\ncalA_{\rm sh}(\nbp_1;\nbz)$ in the azimuth. Then,
 \begin{align}
\label{eq:alpha}
\alpha(\nbp^{(2)};\nbz) &= \max\left(0,\frac{\epsilon(\nbp^{(2)};\nbz)}{\theta(\nbp_2;\nbz)} \right)
\end{align}
where
\begin{align}
\label{eq:epsilon}
~\epsilon(\nbp^{(2)};\nbz) &= \begin{cases}
			 & \min(u(\nbp_1;\nbz),u(\nbp_2;\nbz)) - \max(l(\nbp_1;\nbz),l(\nbp_2;\nbz)), ~\mbox{\em if} \\
			 & \hspace{30mm} l(\nbp_1;\nbz) \leq u(\nbp_1;\nbz), l(\nbp_2;\nbz) \leq u(\nbp_2;\nbz) \\
			 & 2\pi - (\max(l(\nbp_1;\nbz),l(\nbp_2;\nbz)) - \min(u(\nbp_1;\nbz),u(\nbp_2;\nbz))), ~\mbox{\em if} \\
			 & \hspace{30mm} l(\nbp_1;\nbz) > u(\nbp_1;\nbz), l(\nbp_2;\nbz) > u(\nbp_2;\nbz) \\
			 & \max(u(\nbp_2;\nbz)-l(\nbp_1;\nbz),u(\nbp_1;\nbz)-l(\nbp_2;\nbz)),~\mbox{\em else.}
			\end{cases}
 \end{align}
\end{lemma}
\begin{IEEEproof}
 See Appendix \ref{app:alpha}
\end{IEEEproof}

The visible region beyond  a radius $r_2$ can be decomposed into the union of two sets, $\ncalV_{\rm in}(\nbp^{(k)};\nbz)$ and $\ncalV_{\rm out}(\nbp^{(k)};\nbz)$, which are defined as follows:
\begin{align}
 \ncalV_{\rm in}(\nbp^{(k)};\nbz) &= \{\nbp \in \ncalV(\nbz): r > r_2, \phi \in \ncalI(\nbp_1;\nbz) \cup \ncalI(\nbp_2;\nbz) \}\\
 \ncalV_{\rm out}(\nbp^{(k)};\nbz) &= \{\nbp \in \ncalV(\nbz): r > r_2, \phi \notin \ncalI(\nbp_1;\nbz) \cup \ncalI(\nbp_2;\nbz) \}.
\end{align}
$\ncalV_{\rm in}(\nbp^{(k)};\nbz)$ is the (vertically) striped region in Fig. \ref{fig:quasi_setup} and $\ncalV_{\rm out}(\nbp^{(k)};\nbz)$ is a subset of the annular region from $r_2$ to $R$, excluding the azimuthal end points of $\ncalA_{\rm sh}(\nbp_1;\nbz) \cup \ncalA_{\rm sh}(\nbp_2;\nbz)$, i.e., the checkered region in Fig. \ref{fig:quasi_setup}. Using the terminology defined so far, $A_{v}^{(k)}(\nbp^{(k)};\nbz)$ can be expressed as follows:
\begin{align}
\label{eq:Avk}
 A_v^{(k)}(\nbp^{(k)};\nbz) &= A_{n2}(\nbp^{(2)};\nbz)+A_{f}(\nbp^{(k)};\nbz), \\
\label{eq:Anear2} 
 \mbox{where}~ A_{n2}(\nbp^{(2)};\nbz) &\triangleq \pi r_{2}^2 - \left(\frac{\theta(\nbp_1;\tilde{\nbz})}{2}r_{2}^2-\frac{1}{2}x(\nbp_1;\tilde{\nbz})r_{1}\right) \\
 \tilde{\nbz}&= [\lambda_0 ~ L ~ r_2] \\
\label{eq:Afar} 
 A_{f}(\nbp^{(k)};\nbz) &\triangleq \nu_2(\ncalV_{\rm in}(\nbp^{(k)};\nbz)) + \nu_2(\ncalV_{\rm out}(\nbp^{(k)};\nbz)).
\end{align}
In (\ref{eq:Avk})-(\ref{eq:Afar}), $A_{n2}(\nbp^{(2)};\nbz)$ denotes the visible area up to the location of the second nearest obstacle (i.e., the area of the white region in Fig. \ref{fig:quasi_setup}) and $A_{f}(\nbp^{(k)};\nbz)$ denotes the remaining visible area, beyond the second nearest obstacle.

For $k > 2$, evaluating the pairwise shadow region overlaps is, in general, insufficient, as more than two obstacles may contribute to a common overlapping region. Since it is not straightforward to ensure that the areas of all overlapping shadowed regions are counted exactly once, $A_{f}(\nbp^{(k)};\nbz)$ is difficult to compute exactly. Consequently, $f_{A_v(\nbz)}(.)$ is hard to characterize in closed form, as well. Hence, we focus on approximating $A_{f}(\nbp^{(k)};\nbz)$ in the remainder of this section, which shall then be used to derive a tractable approximation for $b(\lambda,\nbz)$ in the following section.

\begin{figure}
 \centering
  \includegraphics[scale=0.7]{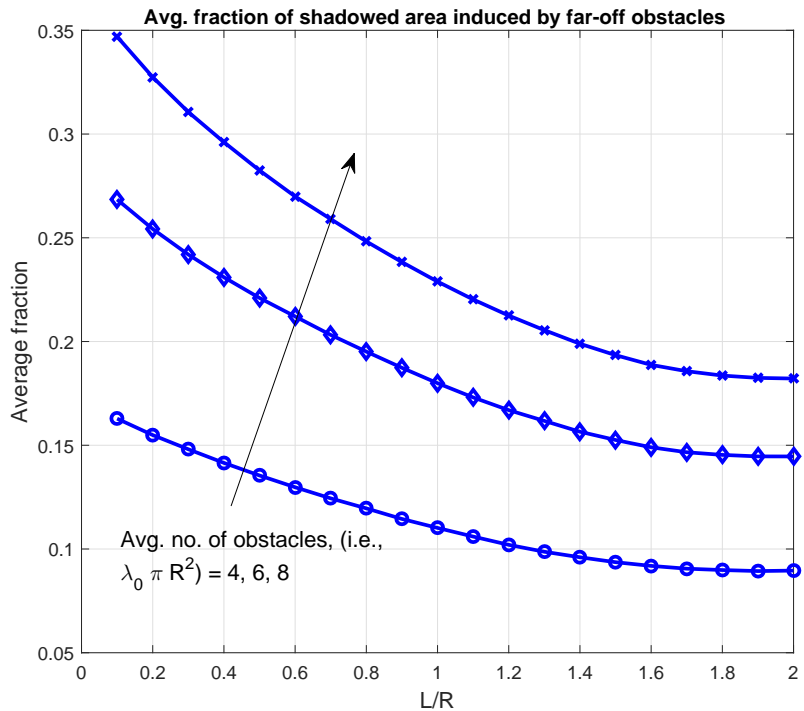}
 \caption{The $y$-axis plots the average fraction of the shadowed area. The curves have been generated by averaging over $10^6$ Monte-Carlo simulations.}
 \label{fig:Ash_frac}
\end{figure}

%\begin{figure}
% \centering
% \includegraphics[scale=0.7]{Ash_frac.eps}
% \caption{The nearest two obstacles are responsible for $65-90\%$ of the shadowed area. The above curves have been generated by averaging over $10^6$ Monte-Carlo simulations.}
% \label{fig:Ash_frac}
%\end{figure}

% 1. Specify that Monte-Carlo simulations have been used to generate these curves.
% 2. Specify that this nearest two-obstacle approximation is justified only for a small number of obstacles.
Since nearer obstacles induce larger shadow regions, it is intuitive that the nearest two obstacles should be responsible for a large fraction of the total shadowed area. To quantify this notion, let $\gamma(\nbz)=\nbbE[A_f(\nbp^{(k)};\nbz)/A_v^{(k)}(\nbp^{(k)};\nbz)]$ denote the average fraction of the shadowed area contributed by the far-off obstacles, where the expectation is over both the number as well as the locations of the obstacles. $\gamma(\nbz)$ is plotted in Fig. \ref{fig:Ash_frac} as a function of the normalized obstacle length, $L/R$, by averaging over $10^6$ Monte-Carlo simulations. Unsurprisingly, $\gamma(\nbz)$ increases with the number of obstacles as there is a greater possibility of a non-overlapping far-off shadow region. However, the likelihood of such an outcome reduces with increasing obstacle size and therefore, $\gamma(\nbz)$ is monotonically decreasing in $L/R$. Hence, when there are a small number of obstacles on average, the nearest two account for most of the shadowed area (in excess of $60\%$, on average, when the average number of obstacles is at most eight, as seen in Fig.~\ref{fig:Ash_frac}). 

Thus, conditioned on $\nbp^{(2)}$, it is reasonable to approximate the shadowed area due to the remaining obstacles by its mean value. In other words, $A_{f}(\nbp^{(k)};\nbz)$ can be approximated by its mean value, conditioned on $\nbp^{(2)}$. We refer to this as the \emph{nearest two-obstacle approximation}, which is formally expressed below:
 
\begin{approximation}[Nearest two-obstacle approximation] \label{approx:near2}For $k \geq 2$ and a small number of obstacles on average\footnote{Based on Fig.~\ref{fig:Ash_frac}, at most eight obstacles on average is a reasonable heuristic.},
 \begin{align}
A_v^{(k)}(\nbp^{(k)};\nbz) &\approx A_v^{(2+)}(\nbp^{(2)};\nbz) \notag \notag\\
\label{eq:Av_assump}
&\triangleq A_{n2}(\nbp^{(2)};\nbz) + \nbbE[A_{f}(\nbp^{(k)};\nbz)|\nbp^{(2)}] \\
\label{eq:Av_assump2}
&\approx A_{n2}(\nbp^{(2)};\nbz) + \nbbE[\nu_2(\ncalV_{\rm out}(\nbp^{(k)};\nbz))|\nbp^{(2)}].
\end{align}
\end{approximation}
In evaluating the conditional mean of $A_{f}(\nbp^{(k)};\nbz)$ in (\ref{eq:Av_assump}), given $\nbp^{(2)}$, we average over both the \emph{number} and the \emph{locations} of the far-off obstacles, i.e., over both $k$ and $\nbp^{(3:k)}$, respectively. The approximation in (\ref{eq:Av_assump2}) is obtained from (\ref{eq:Afar}) by ignoring the term $\nbbE[\nu_2(\ncalV_{\rm in}(\nbp^{(k)});\nbz)|\nbp^{(2)}]$ (i.e., the average area of the striped region in Fig.~\ref{fig:quasi_setup}) for the sake of tractability. However, it is easy to observe from Fig.~\ref{fig:quasi_setup} that the area of the striped region increases with increasing obstacle size. As a result, the approximation in (\ref{eq:Av_assump2}) may not be reasonable beyond a certain value of $L$. In the following lemma, we derive an expression for $\nbbE [\nu_2(\ncalV_{\rm out}(\nbp^{(k)};\nbz))|\nbp^{(2)}]$.

% Comment about this underestimating the {\color{red}visible} area

\begin{lemma} \label{lem:Aindep_2+}
Conditioned on the nearest two obstacles, the average visible area over $\ncalV_{\rm out}$ is given by
\begin{align}
 \nbbE[\nu_2(\ncalV_{\rm out}(\nbp^{(k)};\nbz))|\nbp^{(2)}]&= \left(2\pi - \theta(\nbp_1;\nbz)-(1-\alpha(\nbp^{(2)};\nbz))\theta(\nbp_2;\nbz)\right) \times \notag \\
 & \displaystyle\int\limits_{r_{2}}^R \exp\left(-2\lambda_0 \displaystyle\int\limits_{r_{2}}^r \rho \min\left(\arctan\left(\frac{L}{2\rho}\right), \arccos\left(\frac{\rho}{r}\right)\right)~ {\rm d}\rho \right) {\rm d}r.
\end{align}
\end{lemma}
\begin{IEEEproof}
 See Appendix \ref{app:Aindep_2+}.
\end{IEEEproof}
\begin{remark}
% Quasi-independent blocking 
The nearest two-obstacle approximation characterizes the visible area beyond the second nearest obstacle only by its mean. However, from Lemma \ref{lem:pbs_indep}, this is equivalent to assuming independent blocking beyond the second nearest obstacle. Hence, the nearest two-obstacle approximation can also be interpreted as a `quasi-independent blocking assumption'.
\end{remark}
In the next section, we derive a tractable approximation for $b(\lambda,\nbz)$ using the nearest two-obstacle approximation.

\section{A Tractable Approximation for $b(\lambda,\nbz)$}
\label{sec:tractable_approx}
%In the next section, we show that the nearest two-obstacle approximation is equivalent to assuming \emph{quasi-independent} blocking. 
 
%\subsection{Quasi-independent blocking}
Let $b_k(\lambda,\nbz)$ denote the blind spot probability, conditioned on $k$ obstacles being present, for anchor intensity $\lambda$ and parameter vector $\nbz$. By first conditioning and then averaging over the obstacle locations, $b_k(\lambda,\nbz)$ can be expressed as follows:
\begin{align}
\label{eq:pbs_k_obst1}
 b_k(\lambda,\nbz) &=\displaystyle\int\limits_0^{2\pi} \displaystyle\int\limits_{0}^R f^{(k)}(\nbp_1) {\rm d}\nbp_1 \cdots \displaystyle\int\limits_0^{2\pi} \displaystyle\int\limits_{r_{k-1}}^R g(A_v^{(k)}(\nbp^{(k)};\nbz);\lambda) f^{(k)}(\nbp_k|\nbp^{(k-1)}) {\rm d}\nbp_k\\
 \label{eq:pbs_k_obst2}
&= \displaystyle\int\limits_0^{2\pi} \displaystyle\int\limits_{0}^R f^{(k)}(\nbp_1) {\rm d}\nbp_1 \cdots \displaystyle\int\limits_0^{2\pi} \displaystyle\int\limits_{r_{k-1}}^R g(A_v^{(k)}(\nbp^{(k)};\nbz);\lambda) f^{(k)}(\nbp_k|\nbp_{k-1}) {\rm d}\nbp_k, 
\end{align}
where ${\rm d}\nbp_i=r_i {\rm d}r_i {\rm d}\phi_i$ ($i=1,\cdots, k$) and $f^{(k)}(\nbp_i|\nbp^{(i-1)})$ in (\ref{eq:pbs_k_obst1}) denotes the conditional pdf of the location of the $i$-th $(2 \leq i \leq k)$ nearest obstacle given the location(s) of the other obstacles that are closer to the target than it, when a total of $k$ obstacles are present. Similarly, $f^{(k)}(\nbp_1)$ denotes the pdf of the location of the nearest obstacle. The simplification in (\ref{eq:pbs_k_obst2}) is a result of the Markov property, since $r_i$ lies in the interval $[r_{i-1},R]$ and is therefore independent of $r_j$ for $j\in \{1, \cdots, i-2\}$, given $r_{i-1}$.   

For $k<2$, $b_k(\lambda,\nbz)$ is expressed as follows:
\begin{align}
\label{eq:pbs_0}
 b_0(\lambda,\nbz)&= g(A_v^{(0)}(\varnothing;\nbz);\lambda) \\
\label{eq:pbs_1} 
 b_1(\lambda,\nbz)&= \displaystyle\int\limits_0^{2\pi} \displaystyle\int\limits_0^R g(A_v^{(1)}(\nbp_1;\nbz);\lambda) \frac{1}{\pi R^2} {\rm d}\nbp_1.
\end{align}

For $k\geq 2$, the nearest two-obstacle approximation is used to simplify $b_k(\lambda,\nbz)$, as given below,
\begin{align}
\label{eq:b_k}
 b_k(\lambda,\nbz)&\approx b_k^{(2+)} (\lambda,\nbz)\\
 &\triangleq \displaystyle\int\limits_0^{2\pi} \displaystyle\int\limits_{0}^R f^{(k)}(\nbp_1) {\rm d}\nbp_1  \displaystyle\int\limits_0^{2\pi} \displaystyle\int\limits_{r_1}^R f^{(k)}(\nbp_2|\nbp_1) {\rm d}\nbp_2 \cdots \displaystyle\int\limits_0^{2\pi} \displaystyle\int\limits_{r_{k-1}}^R g(A_v^{(2+)}(\nbp^{(2)};\nbz);\lambda) f^{(k)}(\nbp_{k}|\nbp_{k-1}) {\rm d}\nbp_k \\
 &= \displaystyle\int\limits_0^{2\pi} \displaystyle\int\limits_{0}^R f^{(k)}(\nbp_1) {\rm d}\nbp_1 \displaystyle\int\limits_0^{2\pi} \displaystyle\int\limits_{r_{1}}^R g(A_v^{(2+)}(\nbp^{(2)};\nbz);\lambda) f^{(k)}(\nbp_2|\nbp_1) {\rm d}\nbp_2,
\end{align}
where $A_v^{(2+)}(\nbp^{(2)};\nbz)$ is given by (\ref{eq:Av_assump2}). The expressions for $f^{(k)}(\nbp_1)$ and $f^{(k)}(\nbp_2|\nbp_1)$ are as follows:
 \begin{align}
 \label{eq:f1}
  f^{(k)}(\nbp_1) &= \frac{k}{\pi R^2} \left(\frac{R^2-r_1^2}{R^2}\right)^{k-1} \\
  \label{eq:f2|1}
  f^{(k)}(\nbp_2|\nbp_1) &= \frac{k-1}{\pi(R^2-r_1^2)}\left(\frac{R^2-r_2^2}{R^2-r_1^2}\right)^{k-2}
 \end{align}
with (\ref{eq:f1}) and (\ref{eq:f2|1}) following as a result of the $k$ obstacle mid-points being independently and uniformly distributed over $\ncalD_\nbo(R)$.

Using (\ref{eq:pbs_0})-(\ref{eq:f2|1}), an approximate expression for $b(\lambda,\nbz)$ can be derived by first conditioning and then averaging over the number of obstacles, $k$, in the following manner:
\begin{align}
\label{eq:pbs_exp}
 b(\lambda,\nbz) &= \displaystyle\sum\limits_{k=0}^{\infty} b_k(\lambda,\nbz) e^{-\lambda_0 \pi R^2} \frac{(\lambda_0 \pi R^2)^k}{k!} \\
\label{eq:pbs_approx2} 
 &\approx b_0(\lambda,\nbz)e^{-\lambda_0 \pi R^2} + b_1(\lambda,\nbz)e^{-\lambda_0 \pi R^2}(\lambda_0 \pi R^2) + \displaystyle\sum\limits_{k=2}^{\infty} b_k^{(2+)}(\lambda,\nbz) e^{-\lambda_0 \pi R^2} \frac{(\lambda_0 \pi R^2)^k}{k!} \\
\label{eq:pbs_2+_penult}
&= g(A_v^{(0)}(\varnothing;\nbz);\lambda) e^{-\lambda_0 \pi R^2} + \left(\displaystyle\int\limits_0^{2\pi} \displaystyle\int\limits_0^R g(A_v^{(1)}(\nbp_1;\nbz);\lambda) \frac{1}{\pi R^2} r_{1} {\rm d}r_{1} {\rm d}\phi_{1}\right)e^{-\lambda_0 \pi R^2} (\lambda_0 \pi R^2) \notag \\
&+ \displaystyle\int\limits_0^{2\pi} \displaystyle\int\limits_0^R {\rm d}\nbp_1 \displaystyle\int\limits_0^{2\pi}  \displaystyle\int\limits_{r_1}^R g(A_v^{(2+)}(\nbp^{(2)};\nbz);\lambda) e^{-\lambda_0 \pi R^2}\left(\displaystyle\sum\limits_{k=2}^\infty f^{(k)}(\nbp_1) f^{(k)}(\nbp_2|\nbp_1) \frac{(\lambda_0 \pi R^2)^k}{k!}\right) {\rm d}\nbp_2 \\
\label{eq:pbs_2+_final}
&= g(A_v^{(0)}(\varnothing;\nbz);\lambda) e^{-\lambda_0 \pi R^2} + \left(\displaystyle\int\limits_0^{2\pi} \displaystyle\int\limits_0^R g(A_v^{(1)}(\nbp_1;\nbz);\lambda) \frac{1}{\pi R^2} r_{1} {\rm d}r_{1} {\rm d}\phi_{1}\right)e^{-\lambda_0 \pi R^2} (\lambda_0 \pi R^2) \notag \\
&+ \displaystyle\int\limits_0^{2\pi} \displaystyle\int\limits_0^R r_1{\rm d}r_1 {\rm d}\phi_1 \displaystyle\int\limits_0^{2\pi}  \displaystyle\int\limits_{r_1}^R g(A_v^{(2+)}(\nbp^{(2)};\nbz);\lambda) \lambda_0^2 e^{-\lambda_0 \pi r_2^2} ~r_2 {\rm d}r_2 {\rm d}\phi_2 \\
\label{eq:}
&\triangleq b^{(2+)}(\lambda,\nbz).
\end{align}
For all practical purposes, the average number of obstacles is rarely less than two. Hence, the third in the summation in (\ref{eq:pbs_2+_final}) is the most significant. We now proceed to determine the conditions under which $b^{(2+)}(\lambda,\nbz)$ is a \emph{good} approximation for $b(\lambda,\nbz)$.
%\begin{figure}
% \centering
%  \includegraphics[scale=0.6]{cA_fig}
%  \caption{For $0 \leq c(A_{n2}(\nbz)) < 2/\lambda$, the monotonicity of $c(A_{n2}(\nbz))$ is intuitive from Fig. \ref{fig:Jensen_proof}, as $A_{n2}(\nbz)+c(A_{n2}(\nbz))$ denotes the $x$-coordinate of the point at which the line passing through $(A_{n2}(\nbz),g(A_{n2}(\nbz);\lambda))$ is tangential to $g(\cdot,\lambda)$.}
%  \label{fig:c}
%\end{figure}  
\begin{theorem}
\label{thm:approx_justify_2}
Given $\nbz$, $b^{(2+)}(\lambda,\nbz) \geq b^{\rm ind}(\lambda,\nbz)$ over $\{(\lambda,\nbz): \lambda \nbbE[A_v(\nbz)|\mathsf{K}_2] \geq 3.3836 \}$, where $\mathsf{K}_2$ denotes the event that there are at least two obstacles present in $\ncalD_\nbo(R)$.
\end{theorem}
\begin{IEEEproof}
 By conditioning on $\mathsf{K}_2$ and $\mathsf{K}_2^c$, $\nbbE[A_v(\nbz)]$ can be expressed as follows:
\begin{align}
 \label{eq:Eav_K2expansion}
 \nbbE[A_v(\nbz)] &= \nbbE[A_v(\nbz)|\mathsf{K}_2^c]\nbbP(\mathsf{K}_2^c) + \nbbE[A_v(\nbz)|\mathsf{K}_2]\nbbP(\mathsf{K}_2).
\end{align}
Clearly, $\nbbE[A_v(\nbz)|\mathsf{K}_2] \leq \nbbE[A_v(\nbz)|\mathsf{K}_2^c]$ as the average visible area can only decrease as the number of obstacles increases. Since $g(x;\lambda)$ is convex if and only if $\lambda x\geq 2$, the following holds, from Jensen's inequality, for $\lambda \nbbE[A_v(\nbz)|\mathsf{K}_2^c] \geq \lambda \nbbE[A_v(\nbz)|\mathsf{K}_2] \geq 2$,
\begin{align}
 \label{eq:ind1}
    b^{\rm ind}(\lambda,\nbz)&= g(\nbbE[A_v(\nbz)];\lambda) \\
   \label{eq:ind2}
   &= g(\nbbE[A_v(\nbz)|\mathsf{K}_2^c] \nbbP(\mathsf{K}_2^c) + \nbbE[A_v(\nbz)|\mathsf{K}_2] \nbbP(\mathsf{K}_2);\lambda) \\
   \label{eq:indJen1}
   &\leq g(\nbbE[A_v(\nbz)|\mathsf{K}_2^c];\lambda) \nbbP(\mathsf{K}_2^c) + g(\nbbE[A_v(\nbz)|\mathsf{K}_2];\lambda) \nbbP(\mathsf{K}_2).
\end{align}
Furthermore, from Theorem \ref{thm:jensens}, we have the following inequality for $\lambda \nbbE[A_v(\nbz)|\mathsf{K}_2^c] \geq 3.3836$,
\begin{align}
 \label{eq:indJen2}
 g(\nbbE[A_v(\nbz)|\mathsf{K}_2^c];\lambda) \leq \nbbE[g(A_v(\nbz);\lambda)|\mathsf{K}_2^c].
\end{align}
By further conditioning $\mathsf{K}_2$ on the obstacle locations, $\nbbE[A_v(\nbz)|\mathsf{K}_2]$ can be expressed as follows:
\begin{align}
 \label{eq:Eav_K2}
 \nbbE[A_v(\nbz)|\mathsf{K}_2] &= \nbbE_{\nbp^{(2)}}[A_{n2}(\nbp^{(2)};\nbz)+\nbbE[A_f(\nbp^{(k)};\nbz)|\nbp^{(2)}]].
\end{align}
\begin{remark}
$\nbbE[A_v(\nbz)|\mathsf{K}_2]$ can also be obtained by averaging the expression in (\ref{eq:Avk}) over $k$. The expression in (\ref{eq:Eav_K2}) is an equivalent representation of the same quantity.
\end{remark}
Again, from Theorem \ref{thm:jensens}, the following inequality holds for $\lambda \nbbE[A_v(\nbz)|\mathsf{K}_2] \geq 3.3836$
\begin{align}
\label{eq:g_K2}
 g(\nbbE[A_v(\nbz)|\mathsf{K}_2];\lambda) &= g(\nbbE_{\nbp^{(2)}}[A_{n2}(\nbp^{(2)};\nbz)+\nbbE[A_f(\nbp^{(k)};\nbz)|\nbp^{(2)}]];\lambda) \\
 \label{eq:indJen3}
 &\leq \nbbE_{\nbp^{(2)}}[g(A_{n2}(\nbp^{(2)};\nbz)+\nbbE[A_f(\nbp^{(k)};\nbz)|\nbp^{(2)}];\lambda)].
\end{align}
Thus, from (\ref{eq:Eav_K2expansion})-(\ref{eq:indJen3}), for $\lambda \nbbE[A_v(\nbz)|\mathsf{K}_2^c] \geq \lambda \nbbE[A_v(\nbz)|\mathsf{K}_2] \geq 3.3836$, we have
\begin{align}
 b^{\rm ind}(\lambda,\nbz) &= g(\nbbE[A_v(\nbz)];\lambda) \notag \\
 &\leq  \nbbE[g(A_v(\nbz);\lambda)|\mathsf{K}_2^c] \nbbP(\mathsf{K}_2^c) + \nbbE_{\nbp^{(2)}}[g(A_{n2}(\nbp^{(2)};\nbz)+\nbbE[A_f(\nbp^{(k)};\nbz)|\nbp^{(2)}];\lambda)] \nbbP(\mathsf{K}_2) \notag \\
 &= b^{(2+)}(\lambda,\nbz).
\end{align}
\end{IEEEproof}
\begin{remark}
Similar to Theorem~\ref{thm:jensens}, Theorem~\ref{thm:approx_justify_2} represents a sufficient, but not necessary, condition. 
\end{remark}

\begin{theorem}
\label{thm:approx_justify_1}
Given $\nbz$ and $\lambda$, $b^{(2+)}(\lambda,\nbz) - b(\lambda,\nbz) \leq c(\lambda;\nbz)$, where $c(\lambda;\nbz) \in (0,1)$ is a decreasing function in $\lambda$.
\end{theorem}
\begin{IEEEproof}
Conditioning on $\mathsf{K}_2$ and $\mathsf{K}_2^c$, $b(\lambda,\nbz)$ and $b^{(2+)}(\lambda,\nbz)$ can be expressed as follows:
\begin{align}
 b(\lambda,\nbz)&= \nbbE[g(A_v(\nbz);\lambda)|\mathsf{K}_2^c]~\nbbP(\mathsf{K}_2^c) + \nbbE[g(A_v(\nbz);\lambda)|\mathsf{K}_2]~\nbbP(\mathsf{K}_2) \notag \\
 \label{eq:true}
 &= \nbbE[g(A_v(\nbz);\lambda)|\mathsf{K}_2^c]\nbbP(\mathsf{K}_2^c) + \nbbE_{\nbp^{(2)}}[\nbbE[g(A_{n2}(\nbp^{(2)};\nbz)+A_{f}(\nbp^{(k)};\nbz);\lambda)|\nbp^{(2)}]] \nbbP(\mathsf{K}_2) \\
 \label{eq:approx}
 b^{(2+)}(\lambda,\nbz)&= \nbbE[g(A_v(\nbz);\lambda)|\mathsf{K}_2^c]\nbbP(\mathsf{K}_2^c) + \nbbE_{\nbp^{(2)}}[g(A_{n2}(\nbp^{(2)};\nbz)+\nbbE[A_{f}(\nbp^{(k)};\nbz)|\nbp^{(2)}];\lambda)] \nbbP(\mathsf{K}_2).
\end{align}
Similar to (\ref{eq:Av_assump}), the conditional expectation in (\ref{eq:true}), given $\nbp^{(2)}$, is over both $k$ and $\nbp^{(3:k)}$. Let
\begin{align}
 \label{eq:g1}
 g_1(\nbp^{(2)};\lambda,\nbz) &= \nbbE[g(A_{n2}(\nbp^{(2)};\nbz)+A_{f}(\nbp^{(k)};\nbz);\lambda)|\nbp^{(2)}] \\
 \label{eq:g2}
 g_2(\nbp^{(2)};\lambda, \nbz) &= g(A_{n2}(\nbp^{(2)};\nbz)+\nbbE[A_{f}(\nbp^{(k)};\nbz)|\nbp^{(2)}];\lambda) \\
 \label{eq:R1}
 \ncalR_1(\lambda;\nbz) &:= \{\nbp^{(2)} \in \ncalD_\nbo(R) \times \ncalD_\nbo(R): g_1(\nbp^{(2)};\lambda,\nbz) \geq g_2(\nbp^{(2)};\lambda, \nbz), ~ r_1 \leq r_2 \} \\
 \label{eq:R2}
 \ncalR_2(\lambda;\nbz) &:= \{\nbp^{(2)} \in \ncalD_\nbo(R) \times \ncalD_\nbo(R): g_1(\nbp^{(2)};\lambda, \nbz) < g_2(\nbp^{(2)};\lambda, \nbz),~ r_1 \leq r_2\} \\
 \label{eq:F1}
 \ncalF_1(\lambda;\nbz) &:=\{\nbp^{(2)} \in \ncalD_\nbo(R) \times \ncalD_\nbo(R): A_{n2}(\nbp^{(2)};\nbz) \geq  2/\lambda, ~ r_1 \leq r_2 \} \\
 \label{eq:F2}
 \ncalF_2(\lambda;\nbz) &:=\{\nbp^{(2)} \in \ncalD_\nbo(R) \times \ncalD_\nbo(R): A_{n2}(\nbp^{(2)};\nbz) <  2/\lambda, ~ r_1 \leq r_2 \}
\end{align}
Since $g(x;\lambda)$ is convex whenever $\lambda x \geq 2$, it follows that $g(\cdot;\lambda)$ is convex over the set of $A_{n2}(\nbp^{(2)};\nbz)$ resulting from $\ncalF_1$. Hence, from Jensen's inequality, $\ncalF_1 \subseteq \ncalR_1$. As a result, $\ncalF_2 \supseteq \ncalR_2$, since $\ncalR_1 \cup \ncalR_2 = \ncalF_1 \cup \ncalF_2$. Hence, from (\ref{eq:true})-(\ref{eq:F2}),
\begin{align}
\label{eq:diff_R1R2}
 b^{(2+)}(\lambda,\nbz) - b(\lambda,\nbz) &= \nbbP(\mathsf{K}_2) \left(\displaystyle\int\limits_{\ncalR_1(\lambda;\nbz)} (g_2(\nbp^{(2)};\lambda,\nbz)-g_1(\nbp^{(2)};\lambda,\nbz)) f(\nbp^{(2)}) {\rm d}\nbp^{(2)} \right. \notag \\
 & \hspace{20mm}\left.+ \displaystyle\int\limits_{\ncalR_2(\lambda;\nbz)} (g_2(\nbp^{(2)};\lambda,\nbz)-g_1(\nbp^{(2)};\lambda,\nbz)) f(\nbp^{(2)}) {\rm d}\nbp^{(2)} \right), \\
\end{align}
where $f(\nbp^{(2)})$ denotes the pdf of $\nbp^{(2)}$. Since the integral over $\ncalR_1(\lambda;\nbz)$ is non-positive, we have  
\begin{align} 
\label{eq:diff_R2} 
b^{(2+)}(\lambda,\nbz) - b(\lambda,\nbz) &\leq \nbbP(\mathsf{K}_2) \displaystyle\int\limits_{\ncalR_2(\lambda;\nbz)} (g_2(\nbp^{(2)};\lambda, \nbz)-g_1(\nbp^{(2)};\lambda, \nbz)) f(\nbp^{(2)}) {\rm d}\nbp^{(2)} \\
 \label{eq:diff_F2}
 &\leq \nbbP(\mathsf{K}_2) \displaystyle\int\limits_{\ncalF_2(\lambda;\nbz)} (g_2(\nbp^{(2)};\lambda,\nbz)-g_1(\nbp^{(2)};\lambda,\nbz)) f(\nbp^{(2)}) {\rm d}\nbp^{(2)} \\
 \label{eq:diff_bd}
 &\leq \nbbP(\mathsf{K}_2) \left(1- \min_{\nbu\in \ncalF_2(\lambda;\nbz)} g_1(\nbu;\lambda)\right) \displaystyle\int\limits_{\ncalF_2(\lambda;\nbz)} f(\nbp^{(2)}) {\rm d}\nbp^{(2)} \\
 \label{eq:c}
 &:= c(\lambda;\nbz) ,
\end{align}
where $c(\lambda;\nbz):=\nbbP(\mathsf{K}_2)\left(1- \displaystyle\min\limits_{\nbu\in \ncalF_2(\lambda;\nbz)}g_1(\nbu;\lambda)\right) \nbbP(\nbp^{(2)} \in \ncalF_2(\lambda;\nbz))$ is non-negative and decreasing in $\lambda$ and is bounded above by one.  
\end{IEEEproof}
\begin{remark}
From Theorems \ref{thm:approx_justify_2} and \ref{thm:approx_justify_1}, $b^{\rm ind}(\lambda,\nbz) \leq b^{(2+)}(\lambda,\nbz) \leq b(\lambda,z) + c(\lambda;\nbz)$, for sufficiently large $\lambda$. It is worth pointing out that this inequality relation makes no assumption on the number of obstacles. This implies that $b^{(2+)}(\lambda,\nbz)$ may be a \emph{relatively} more accurate approximation of $b(\lambda,\nbz)$ than $b^{\rm ind}(\lambda,\nbz)$ as $\lambda$ increases, but its accuracy in \emph{absolute} terms is restricted to when the number of obstacles is small, according to Approximation~\ref{approx:near2}.
\end{remark}

To summarize, it is intuitive that obstacles which are closer to the typical target induce greater blocking correlation, with the extent of correlation decreasing with distance. Hence, by taking into account the impact of correlated blocking due to the nearest two obstacles, $b^{(2+)}(\lambda,\nbz)$ achieves a reasonable trade-off between accuracy and tractability.
%%%%%%%%%%%%%%%%%%%%%%%%%%%
\section{Numerical Results} \label{sec:NumResults}
%%%%%%%%%%%%%%%%%%%%%%%%%%%
%We set $R=10{\rm m}$ throughout and f
We consider an average of eight obstacles throughout (i.e., $\lambda_0 \pi R^2 =8$). For each $(\lambda,\nbz)$, the following cases were evaluated: (i) $b(\lambda,\nbz)$, obtained by averaging over $50000$ Monte-Carlo simulations, (ii) $b^{(2+)}(\lambda,\nbz)$, given by (\ref{eq:pbs_2+_final}), and (iii) $b^{\rm ind}(\lambda, \nbz)$. 

\begin{figure}
 \centering
 \includegraphics[scale=0.75]{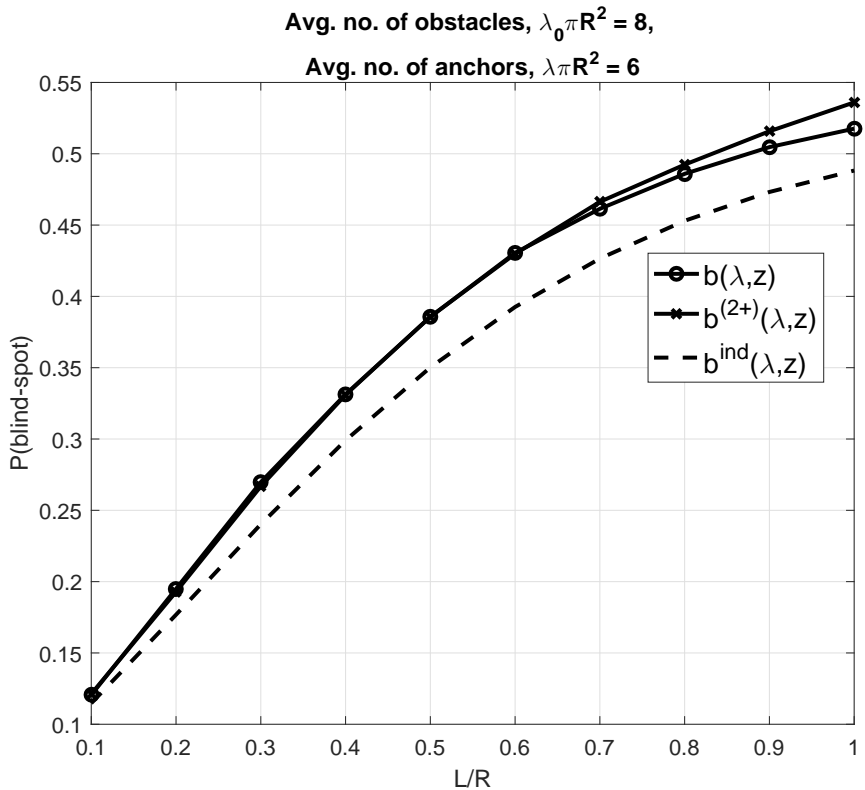}
 \caption{By capturing most of the blocking correlation, $b^{(2+)}(\lambda,\nbz)$ yields an accurate approximation of $b(\lambda,\nbz)$. In contrast, by ignoring the blocking correlation, $b^{\rm ind}(\lambda,\nbz)$ significantly underestimates $b(\lambda,\nbz)$.}
 \label{fig:justify}
\end{figure}

For a fixed average number of anchors, the impact of correlated blocking, which is a function of the normalized obstacle length, $L/R$, on the blind spot probability is shown in Fig. \ref{fig:justify}. For small values of $L/R$ (low blocking correlation), the difference between the three cases is minimal, which is intuitive. However, even for moderate blocking correlation $(L/R=0.5)$, $b^{\rm ind}(\lambda,\nbz)$ significantly underestimates $b(\lambda,\nbz)$. In contrast, $b^{(2+)}(\lambda,\nbz)$ accurately estimates $b(\lambda,\nbz)$ across all levels of blocking correlation. 

\begin{figure}
 \centering
 \begin{subfigure}{0.49\textwidth}
  \centering
  \includegraphics[scale=0.5]{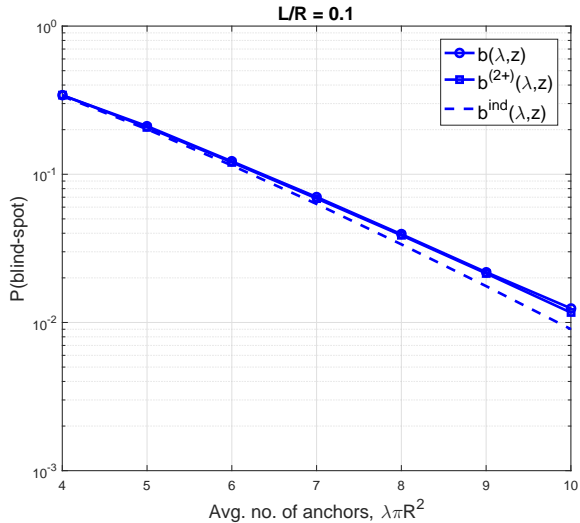}
  \caption{Low blocking correlation: $L/R = 0.1$}
  \label{fig:anchor_low}
 \end{subfigure}
 \hfill
  \begin{subfigure}{0.49\textwidth}
  \centering
  \includegraphics[scale=0.5]{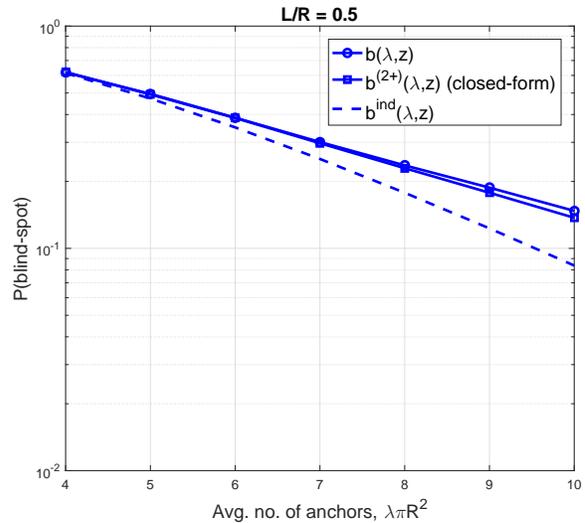}
  \caption{Moderate blocking correlation: $L/R = 0.5$}
  \label{fig:anchor_med}
 \end{subfigure}
 \\
  \begin{subfigure}{0.9\textwidth}
  \centering
  \includegraphics[scale=0.5]{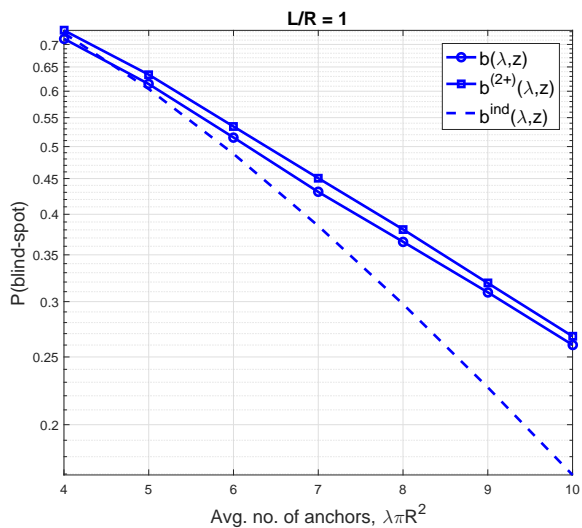}
  \caption{High blocking correlation: $L/R = 1$}
  \label{fig:anchor_high}
 \end{subfigure}
 \hfill 
 \caption{The accuracy of $b^{(2+)}(\lambda,\nbz)$ implies that it can be used to determine the anchor intensity that satisfies $b^{(2+)}(\lambda,\nbz) \approx b(\lambda,\nbz) \leq \mu$, for some threshold, $\mu$.}
 \label{fig:anchor}
\end{figure}

For three different cases of $L/R$, which capture low, moderate and high blocking correlation, the blind spot probability is plotted as a function of the average number of anchors, $\lambda \pi R^2$, in Fig.~\ref{fig:anchor}. For all the cases, $b^{\rm ind}(\lambda,\nbz)$ decreases faster than $b(\lambda,\nbz)$ with increasing $\lambda$, with the rate of divergence being proportional to $L/R$. Since the nearest two-obstacle approximation captures most of the blocking correlation, the rate of decrease of $b^{(2+)}(\lambda,\nbz)$ with respect to $\lambda$ is almost identical to that of $b(\lambda,\nbz)$, leading to a more accurate approximation. Hence, from a design perspective, $b^{(2+)}(\lambda,\nbz)$ can be used to determine $\lambda$ such that $b(\lambda,\nbz) \approx b^{(2+)}(\lambda,\nbz) \leq \mu$. It is worth pointing out that  $b^{(2+)}(\lambda,\nbz) \geq b(\lambda,\nbz)$ for high blocking correlation (\ref{fig:anchor_high}). Although this is consistent with the statement of Theorem~\ref{thm:approx_justify_1}, we believe that the effect of ignoring the term $\nbbE[\nu_2(\ncalV_{\rm in}(\nbp^{(k)});\nbz)|\nbp^{(2)}]$, which is the average area of the striped region in Fig.~\ref{fig:quasi_setup} may also be a contributing factor. As pointed out in Approximation~\ref{approx:near2}, $\nbbE[\nu_2(\ncalV_{\rm in}(\nbp^{(k)});\nbz)|\nbp^{(2)}]$ increases with $L$. Hence, by neglecting its contribution to $A_v^{(2+)}(\nbp^{(2)};\nbz)$ in (\ref{eq:Anear2}), the unshadowed area beyond the second nearest obstacle is systematically underestimated, which may contribute to $b^{(2+)}(\lambda;\nbz)$ being greater than $b(\lambda;\nbz)$.

%The approximation error can be further reduced by considering the blocking correlation beyond second nearest obstacle; however, this comes at the expense of tractability as we would have to deal with more complex overlaps involving the shadowed regions.
%From a design point of view, this graph is the most useful

%%%%%%%%%%%%%%%
\section{Summary}
\label{sec:concl}
In this paper, we set out to analyze the impact of obstacle-induced correlated blocking on the blind spot probability at a typical target location in a localization network. To model the uncertainty in the obstacle locations as well as capture the blocking correlation induced by the obstacle size, we considered a novel stochastic geometry based approach where the obstacles were modeled as random line-segments using a germ-grain model. For anchors deployed according to homogeneous PPP, we characterized the blind spot probability as a function of the pdf of the visible area surrounding a typical target. Furthermore, we showed that the blind spot probability under the independent anchor blocking assumption depends only on the mean visible area, instead of the entire probability distribution, and derived the conditions under which the independent blocking assumption underestimates the true blind spot probability. Since the pdf of the visible area is difficult to characterize in closed form, we derived an approximate expression for the blind spot probability by formulating the \emph{nearest two-obstacle approximation}, which captures the blocking correlation up to the second nearest obstacle and assumes independent blocking due to farther obstacles. This yields a trade-off between accuracy and tractability, wherein our approximation is more accurate than the independent blocking assumption in estimating the true blind spot probability, as the anchor intensity increases. As a result, our approximation provides design insights, such as the intensity with which anchors need to be deployed so that the blind spot probability over the entire region is less than a threshold, $\mu$.

%Finally, our analysis in this chapter presents a solution approach for similar problems involving other obstacle models that can be described as follows:
%\begin{itemize}
% \item [(a)] Model the randomness in the obstacle locations and shapes.
% \item [(b)] Characterize the visible area distribution. 
% \item[(c)] Obtain an (approximate, if necessary) expression for the blind spot probability based on the visible area distribution.
% \item[(d)] Using (c), determine the intensity with which anchors need to be deployed so that the blind spot probability over a region of interest is less than a threshold, $\mu$.
%\end{itemize} 

%%%%%
\appendix
%%%%%
\subsection{Proof of Lemma~\ref{lem:pbs_indep}}
\label{app:pbs_indep}
The visible area, $A_v(\nbz)$, can be expressed as follows:
\begin{align}
 A_v(\nbz) &= \displaystyle\int\limits_0^{2\pi} \displaystyle\int\limits_0^{R} V(\nbp;\nbz)~r{\rm d} r {\rm d}\phi \\
\label{eq:E[A_v]}
\therefore~ \nbbE [A_v(\nbz)] &= \displaystyle\int\limits_0^{2\pi} \displaystyle\int\limits_0^{R} \nbbE [V(\nbp;\nbz)]~r{\rm d} r {\rm d}\phi =\displaystyle\int\limits_0^{2\pi} \displaystyle\int\limits_0^{R} \nbbP (V(\nbp;\nbz)=1)~r{\rm d} r {\rm d}\phi =2\pi \displaystyle\int\limits_0^{R} \nbbP (V(\nbp;\nbz)=1)~r{\rm d} r,
\end{align}
where (\ref{eq:E[A_v]}) follows from the radial symmetry of the system model considered in Section \ref{sec:SysMod}.

For independent anchor blocking, the unblocked anchors can be viewed as a point process obtained by independently sampling the anchor PPP, where the sampling probability of an anchor at $\nbp \in \ncalD_\nbo(R)$, with respect to the typical target, equals $\nbbP(V(\nbp;\nbz)=1)$. As a result, the unblocked anchors to the typical target form a non-homogeneous PPP whose intensity, $\lambda_{\rm ind}(\nbp;\nbz)$, is given by
\begin{align}
 \label{eq:indep_intensity}
 \lambda^{\rm ind}(\nbp;\nbz)= \lambda \mathbb{P}(V(\nbp;\nbz)=1).
\end{align}
For a non-homogeneous anchor PPP with intensity $\lambda^{\rm ind}(\nbp;{\nbz})$, the number of anchors over a circle of radius $R$ has a Poisson distribution with mean $\Lambda(\lambda,\nbz)$, given by
\begin{align}
\label{eq:Lam_simple}
 \Lambda(\lambda,\nbz) = \displaystyle\int\limits_0^{2\pi} \displaystyle\int\limits_0^R \lambda^{\rm ind}(\nbp;\nbz) r {\rm d}r {\rm d}\phi = \lambda \nbbE[A_v(\nbz)], 
\end{align}
where (\ref{eq:Lam_simple}) is obtained from (\ref{eq:E[A_v]}). Hence, the blind spot probability due to independent anchor blocking is given by
\begin{align}
 \label{eq:probBS_indep_pf}
 b^{\rm ind}(\lambda,\nbz)&= e^{-\Lambda(\lambda,\nbz)}\left(1+\Lambda(\lambda,\nbz)+\frac{(\Lambda(\lambda,\nbz))^2}{2}\right) \notag\\
&= e^{-\lambda \nbbE[A_v(\nbz)]}\left(1+\lambda \nbbE[A_v(\nbz)]+\frac{(\lambda \nbbE[A_v(\nbz)])^2}{2}\right) \notag\\
&= g(\nbbE[A_v(\nbz)];\lambda).
\end{align}
\hfill
\IEEEQED

\subsection{Proof of Lemma~\ref{lem:Ash_firstmom}} \label{app:Ash_firstmom}
\begin{figure}
 \centering
 \includegraphics[scale=0.6]{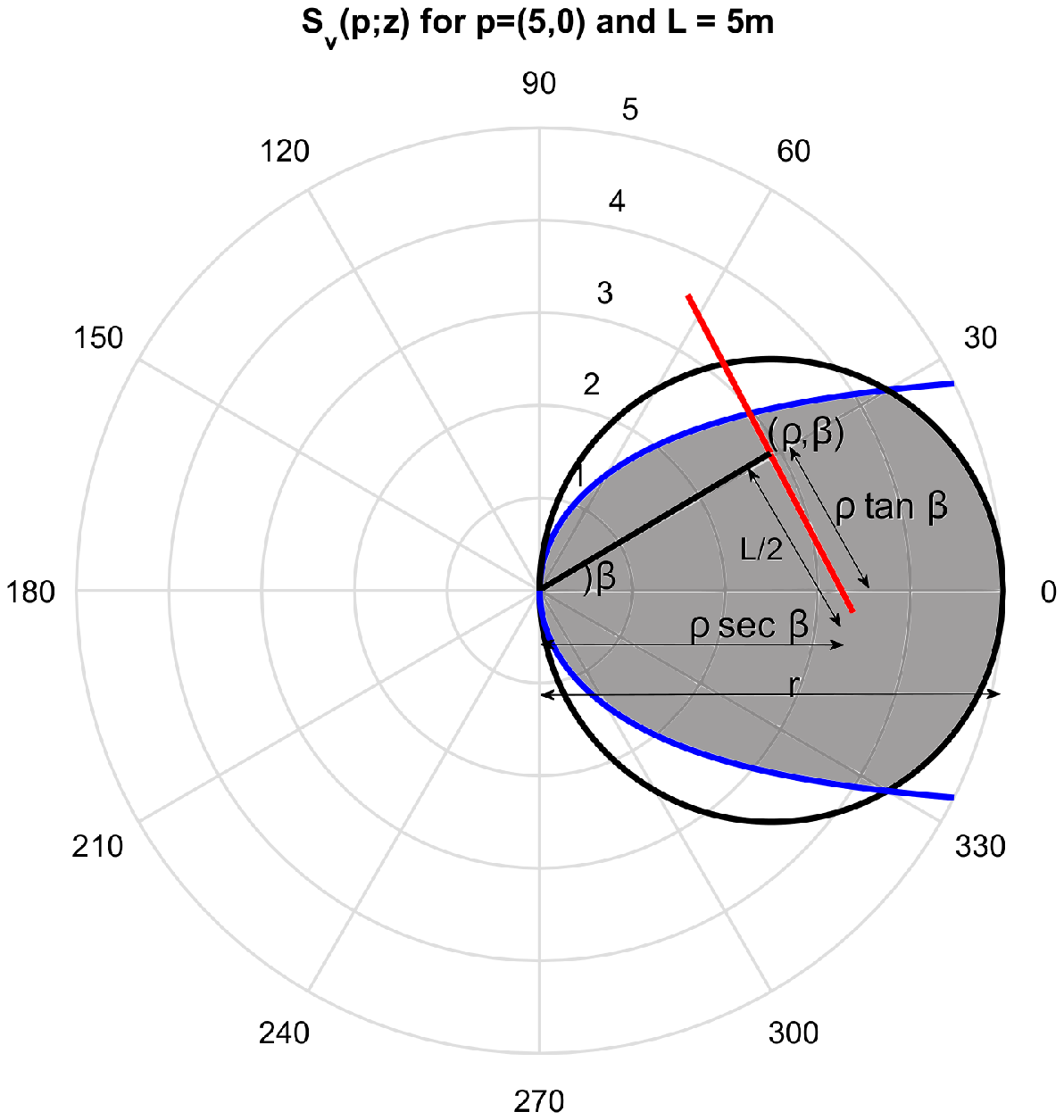}
 \caption{For $\nbp=(5,0)$, the region enclosed by the blue curve corresponds to $S_{V_1}(\nbp;\nbz)=\{(\rho,\beta) \in \nbbR^2 : 0 \leq \rho \tan |\beta-\phi| \leq L/2 \}$. Similarly, the region enclosed by the black curve corresponds to $S_{V_2}(\nbp;\nbz) = \{(\rho,\beta) \in \nbbR^2: 0 \leq \rho \sec |\beta-\phi| \leq r\}$. Hence, the LoS path to $\nbo$ is unblocked if and only if there is no obstacle mid-point in the shaded region, which corresponds to $S_V(\nbp;\nbz)=S_{V_1}(\nbp;\nbz) \cap S_{V_2}(\nbp;\nbz)$.}
 \label{fig:Sv_new}
\end{figure}

An obstacle with mid-point at $(\rho,\beta)$ (in polar coordinates) can block the LoS path between the typical target and an anchor at $\nbp\in \ncalD_\nbo(R)$ if and only if the following conditions are satisfied (see Fig.~\ref{fig:Sv_new})
\begin{align}
 0 \leq \rho \tan |\beta-\phi| \leq L/2 \\
 0 \leq \rho \sec |\beta-\phi| \leq r.
\end{align}
Hence, $V(\nbp;\nbz)=1$ is unblocked if and only if there are no obstacle mid-points in the set $S_V(\nbp;\nbz)= S_{V_1}(\nbp;\nbz) \bigcap S_{V_2}(\nbp;\nbz)$ (Fig.~\ref{fig:Sv_new}), where
\begin{align}
 \label{eq:Sv1}
 S_{V_1}(\nbp;\nbz) &= \{(\rho,\beta) \in \nbbR^2 : 0 \leq \rho \tan |\beta-\phi| \leq L/2 \} \\
 \label{eq:Sv2}
 S_{V_2}(\nbp;\nbz) &= \{(\rho,\beta) \in \nbbR^2 : 0 \leq \rho \sec |\beta-\phi| \leq r\}.
\end{align}
From (\ref{eq:Sv1}) and (\ref{eq:Sv2}), the azimuthal end-points of $S_V(\nbp;\nbz)$ at a radial distance $\rho \in [0,r]$ are given by $\phi \pm \min\left(\arctan\left(\frac{L}{2\rho}\right),\arccos\left(\frac{\rho}{r}\right)\right)$. Therefore,
\begin{align}
 \label{eq:prob_fm}
 \mathbb{P}(V(\nbp;\nbz)=1)&= \nbbP(\mbox{no obstacle mid-point in $S_V(\nbp;\nbz)$}) =e^{-\lambda_0 \nu_2(S_V(\nbp;\nbz))} \\
  \label{eq:area_measure}
\mbox{where}~ \nu_2(S_V(\nbp;\nbz))&= \displaystyle\int\limits_0^r \displaystyle\int\limits_{\phi - \min(\arctan(L/(2\rho)),\arccos(\rho/r))}^{\phi + \min(\arctan(L/(2\rho)),\arccos(\rho/r))} \rho {\rm d}\phi {\rm d}\rho \notag \\
&= 2\displaystyle\int\limits_0^r \rho \min(\arctan(L/(2\rho)),\arccos(\rho/r)) {\rm d}\rho. 
\end{align} 
Substituting (\ref{eq:prob_fm}) and (\ref{eq:area_measure}) in (\ref{eq:E[A_v]}) completes the proof.
 \hfill 
\IEEEQED

\subsection{Proof of Lemma~\ref{lem:alpha}} \label{app:alpha}
\begin{figure}
 \centering
 \begin{subfigure}{0.45\textwidth}
  \includegraphics[scale=0.3]{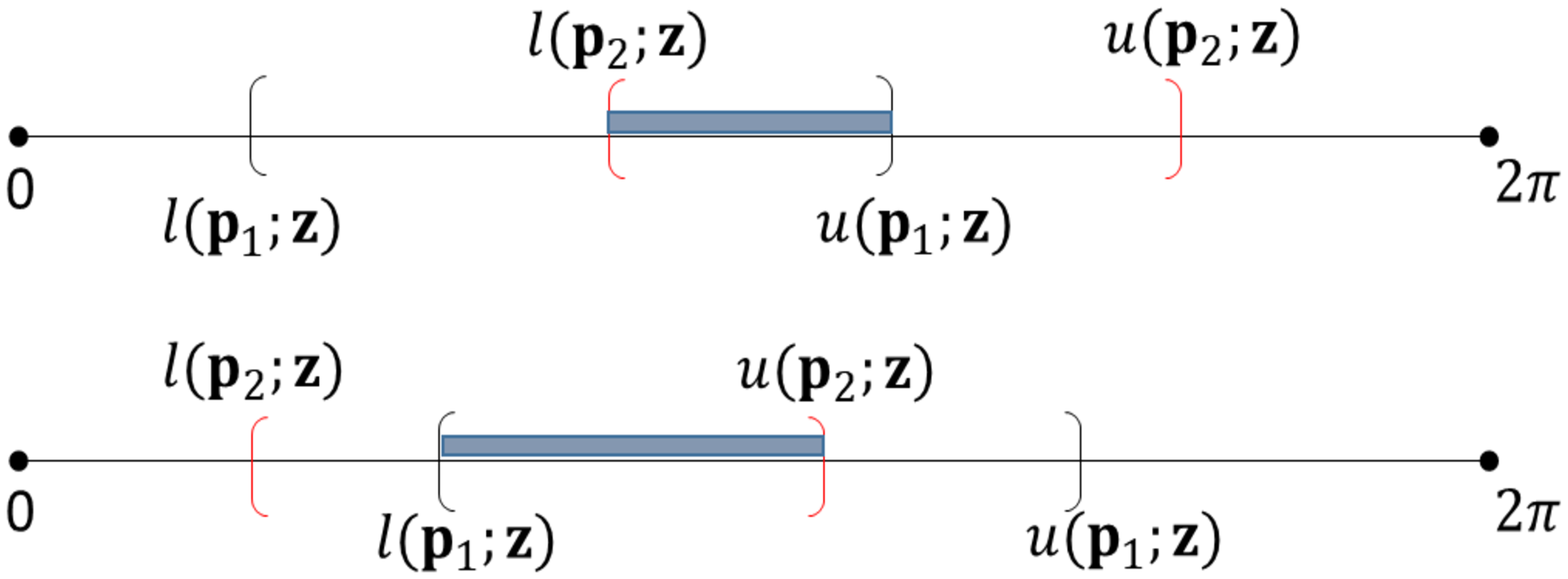}
  \caption{Case 1}
  \label{fig:case_a}
 \end{subfigure}
 ~ 
 \begin{subfigure}{0.45\textwidth}
  \includegraphics[scale=0.3]{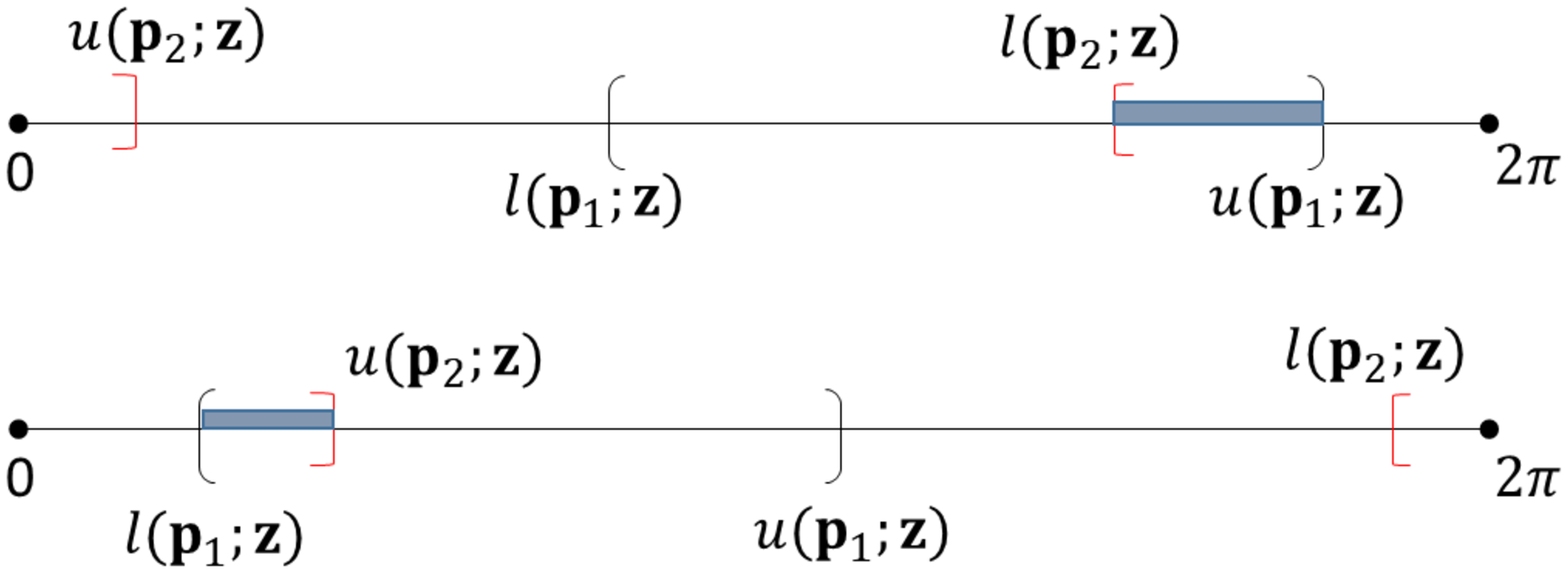}
  \caption{Case 2}
  \label{fig:case_b}
 \end{subfigure}
 ~
 \begin{subfigure}{0.45\textwidth}
  \includegraphics[scale=0.3]{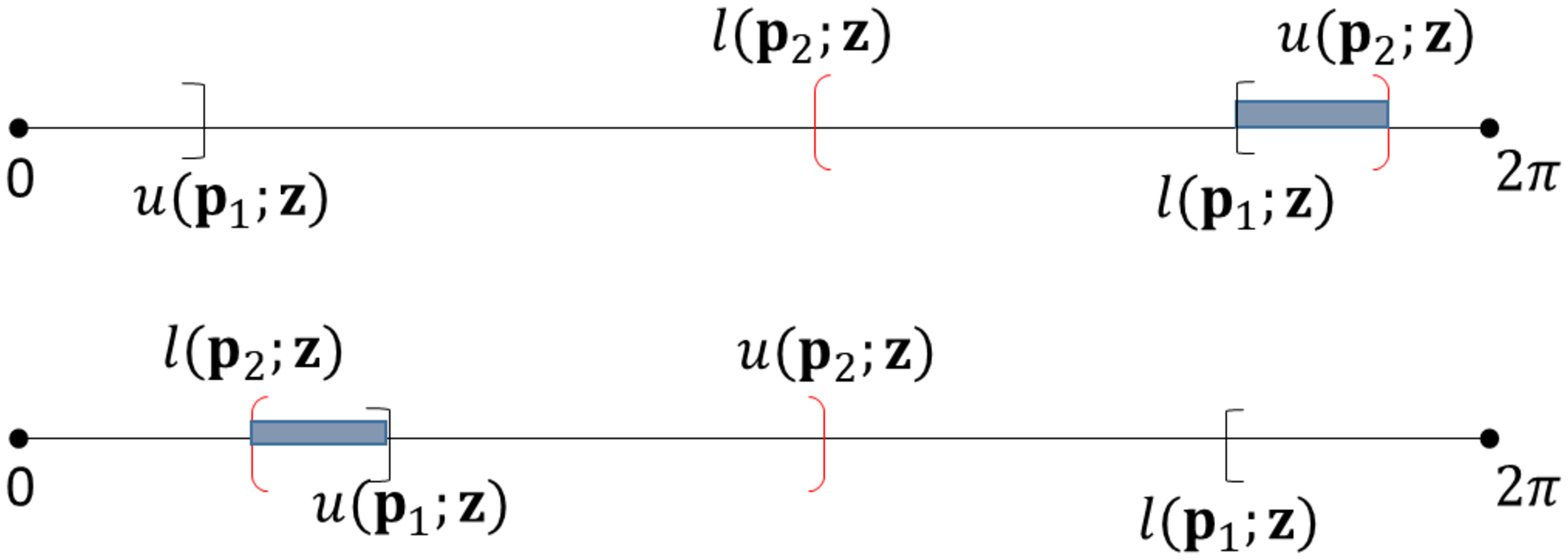}
  \caption{Case 3}
  \label{fig:case_c}
 \end{subfigure}
 ~
 \begin{subfigure}{0.45\textwidth}
  \includegraphics[scale=0.3]{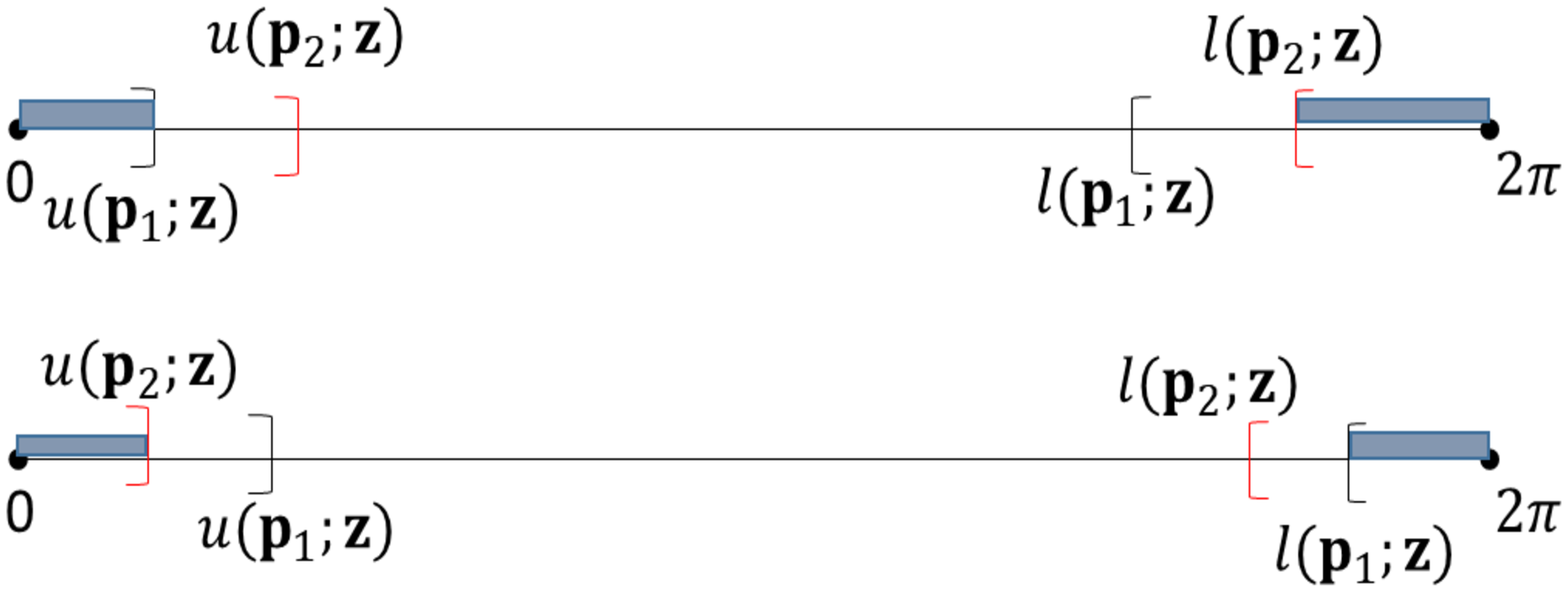}
  \caption{Case 4}
  \label{fig:case_d}
 \end{subfigure}
\caption{Feasible overlap situations between $\ncalA_{\rm sh}(\nbp_1;\nbz)$ and $\ncalA_{\rm sh}(\nbp_2;\nbz)$ in the azimuth coordinate. The size of the interval shaded grey denotes $\epsilon(\nbp^{(2)};\nbz)$.} 
\label{fig:overlap_cases}
\end{figure}
Suppose there exists an overlap between $\ncalA_{\rm sh}(\nbp_2;\nbz)$ and $\ncalA_{\rm sh}(\nbp_1;\nbz)$. Then, let $\epsilon(\nbp^{(2)};\nbz)$ denote the azimuthal width of $\ncalA_{\rm sh}(\nbp_2;\nbz) \cap \ncalA_{\rm sh}(\nbp_1;\nbz)$. The feasible scenarios for the end-points of the intervals $\ncalI(\nbp_1;\nbz)$ and $\ncalI(\nbp_2;\nbz)$ are as follows:
\begin{itemize}
 \item Case 1: $l(\nbp_1;\nbz) \leq u(\nbp_1;\nbz)$ and $l(\nbp_2;\nbz)\leq u(\nbp_2;\nbz)$ \\
 This corresponds to when $\{(r,0):0\leq r \leq R\} \notin \ncalA_{\rm sh}(\nbp_1;\nbz) \cup \ncalA_{\rm sh}(\nbp_2;\nbz)$ (Fig.~\ref{fig:case_a}). For an overlap to occur between $\ncalA_{\rm sh}(\nbp_2;\nbz)$ and $\ncalA_{\rm sh}(\nbp_1;\nbz)$, one of the following conditions must be satisfied:
 \begin{itemize}
  \item[a)] $l(\nbp_1;\nbz) \leq l(\nbp_2;\nbz) < u(\nbp_1;\nbz)$ (top, Fig.~\ref{fig:case_a}).
  \item[b)] $l(\nbp_2;\nbz) \leq l(\nbp_1;\nbz) < u(\nbp_2;\nbz)$ (bottom, Fig.~\ref{fig:case_a}). 
 \end{itemize}
Hence, 
\begin{align}
 \label{eq:case_1}
 \epsilon(\nbp^{(2)};\nbz) &= \min(u(\nbp_1;\nbz),u(\nbp_2;\nbz))-\max(l(\nbp_1;\nbz),l(\nbp_2;\nbz)).
\end{align}
 
 \item Case 2: $l(\nbp_1;\nbz) \leq u(\nbp_1;\nbz)$ and $l(\nbp_2;\nbz)>u(\nbp_2;\nbz)$
This corresponds to when $\{(r,0):0\leq r \leq R\} \notin \ncalA_{\rm sh}(\nbp_1;\nbz)$ and $\{(r,0):0\leq r \leq R\} \in\ncalA_{\rm sh}(\nbp_2;\nbz)$ (Fig.~\ref{fig:case_b}). For an overlap to occur, exactly one of the following conditions must be satisfied: 
\begin{itemize}
  \item[a)] $u(\nbp_2;\nbz) < l(\nbp_1;\nbz) < l(\nbp_2;\nbz) \leq u(\nbp_1;\nbz)$ (top, Fig.\ref{fig:case_b}).
  \item[b)] $u(\nbp_2;\nbz) > l(\nbp_1;\nbz)$ (bottom, Fig.\ref{fig:case_b}).
 \end{itemize}
Hence, 
\begin{align}
 \label{eq:case_2}
 \epsilon(\nbp^{(2)};\nbz) &= \max(u(\nbp_2;\nbz)-l(\nbp_1;\nbz),u(\nbp_1;\nbz)-l(\nbp_2;\nbz)).
\end{align} 

 \item Case 3: $l(\nbp_1;\nbz) > u(\nbp_1;\nbz)$ and $l(\nbp_2;\nbz) \leq u(\nbp_2;\nbz)$
This corresponds to when $\{(r,0):0\leq r \leq R\} \in \ncalA_{\rm sh}(\nbp_1;\nbz)$ and $\{(r,0):0\leq r \leq R\} \notin\ncalA_{\rm sh}(\nbp_2;\nbz)$ (Fig.~\ref{fig:case_c}). For an overlap to occur, exactly one of the following conditions must be satisfied: 
\begin{itemize}
  \item[a)] $l(\nbp_2;\nbz) < u(\nbp_1;\nbz)$ (top, Fig.~\ref{fig:case_c}).
  \item[b)] $u(\nbp_2;\nbz) > l(\nbp_1;\nbz)$ (bottom, Fig.~\ref{fig:case_c}).
 \end{itemize}
Hence, 
\begin{align}
\label{eq:case_3}
 \epsilon(\nbp^{(2)};\nbz) &= \max(u(\nbp_2;\nbz)-l(\nbp_1;\nbz),u(\nbp_1;\nbz)-l(\nbp_2;\nbz)).
\end{align}
 
 \item Case 4: $l(\nbp_1;\nbz) > u(\nbp_1;\nbz)$ and $l(\nbp_2;\nbz) > u(\nbp_2;\nbz)$
This corresponds to when $\{(r,0):0\leq r \leq R\} \in \ncalA_{\rm sh}(\nbp_1;\nbz) \cap \ncalA_{\rm sh}(\nbp_2;\nbz)$ (Fig.~\ref{fig:case_d}). For an overlap to occur, one of the following conditions must be satisfied: 
\begin{itemize}
  \item[a)] $l(\nbp_1;\nbz) < l(\nbp_2;\nbz)$ (top, Fig.~\ref{fig:case_d}).
  \item[b)] $u(\nbp_1;\nbz) > u(\nbp_2;\nbz)$ (bottom, Fig.~\ref{fig:case_d})
 \end{itemize}
Hence, 
\begin{align}
\label{eq:case_4}
 \epsilon(\nbp^{(2)};\nbz) &= 2\pi-(\max(l(\nbp_1;\nbz),l(\nbp_2;\nbz))-\min(u(\nbp_1;\nbz),u(\nbp_2;\nbz))).
\end{align} 
\end{itemize}
From the expressions in (\ref{eq:case_1})-(\ref{eq:case_4}), it is easily seen that $\epsilon(\nbp^{(2)};\nbz)$ is negative if $\ncalA_{\rm sh}(\nbp_1;\nbz) \cap \ncalA_{\rm sh}(\nbp_1;\nbz) = \varnothing$. Hence, from the above cases, the fraction of $\ncalA_{\rm sh}(\nbp_2;\nbz)$ that overlaps with $\ncalA_{\rm sh}(\nbp_1;\nbz)$ is given by,
 \begin{align}
\label{eq:alpha_pf}
\alpha(\nbp^{(2)};\nbz)&= \max\left(0,\frac{\epsilon(\nbp^{(2)};\nbz)}{\theta(\nbp_2;\nbz)}\right), \\
\label{eq:epsilon_pf}
\mbox{where}~\epsilon(\nbp^{(2)};\nbz) &= \begin{cases}
			 & \min(u(\nbp_1;\nbz),u(\nbp_2;\nbz)) - \max(l(\nbp_1;\nbz),l(\nbp_2;\nbz)),~\mbox{if} \\
			 & \hspace{30mm} l(\nbp_1;\nbz) \leq u(\nbp_1;\nbz), l(\nbp_2;\nbz) \leq u(\nbp_2;\nbz) \\
			 & 2\pi - (\max(l(\nbp_1;\nbz),l(\nbp_2;\nbz)) - \min(u(\nbp_1;\nbz),u(\nbp_2;\nbz))),~\mbox{if} \\
			 &\hspace{30mm} l(\nbp_1;\nbz) > u(\nbp_1;\nbz), l(\nbp_2;\nbz) > u(\nbp_2;\nbz) \\
			 & \max(u(\nbp_2;\nbz)-l(\nbp_1;\nbz),u(\nbp_1;\nbz)-l(\nbp_2;\nbz)),\hspace{1mm} \mbox{else.}
			\end{cases}
 \end{align}
\hfill
\IEEEQED

\subsection{Proof of Lemma~\ref{lem:Aindep_2+}} \label{app:Aindep_2+}
Let $\ncalA_{\rm out}(\nbp^{(2)};\nbz)= \{(r,\phi)\in \ncalD_\nbo(R): r>r_2, \phi \notin \ncalI(\nbp_1;\nbz)\cup \ncalI(\nbp_2;\nbz) \}\supseteq \ncalV_{\rm out}(\nbp^{(k)};\nbz)$. Similar to (\ref{eq:E[A_v]}), we have
\begin{align}
 \label{eq:E[Av]_quasi_pf}
 \nbbE[\nu_2(\ncalV_{\rm out}(\nbp^{(k)};\nbz))|\nbp^{(2)}] &= \displaystyle\int\limits_{\nbp\in \ncalA_{\rm out}(\nbp^{(2)};\nbz)} \nbbP (V(\nbp;\nbz)=1) r{\rm d}r {\rm d}\phi.
\end{align}
Due to radial symmetry, the integral in (\ref{eq:E[Av]_quasi_pf}) does not depend on azimuthal coordinate, $\phi$. Hence, 
\begin{align}
 \label{eq:E[Av]_quasi_exp}
 \nbbE[\nu_2(\ncalV_{\rm out}(\nbp^{(k)};\nbz))|\nbp^{(2)}] &=  \varphi_{\rm span}(\ncalA_{\rm out}(\nbp^{(2)};\nbz)) \displaystyle\int\limits_{r_{2}}^R \nbbP (V(\nbp;\nbz)=1) r{\rm d}r, \\
 \label{eq:phi_span}
\mbox{where}~ \varphi_{\rm span}(\ncalA_{\rm out}(\nbp^{(2)};\nbz))&=   2\pi - \theta(\nbp_1;\nbz)-(1-\alpha(\nbp^{(2)};\nbz))\theta(\nbp_2;\nbz). 
\end{align}
In (\ref{eq:phi_span}), $\varphi_{\rm span}(\ncalA_{\rm out}(\nbp^{(2)};\nbz))$  denotes the azimuthal width of $\ncalA_{\rm out}(\nbp^{(2)};\nbz)$. Substituting (\ref{eq:prob_fm}) and (\ref{eq:area_measure}) in (\ref{eq:E[Av]_quasi_pf}), we get the desired result.
\hfill
\IEEEQED

\bibliographystyle{IEEEtran}
\bibliography{IEEEabrv,JunyangBib}
\end{document}